\documentclass[a4paper,10pt]{article}
\newcommand{\proof}{\noindent {\bf Proof: }}

\newtheorem{theorem}{Theorem}

\newtheorem{lemma}{Lemma}
\newtheorem{defi}{Definition}

\def\qed{\hfill $\Box$}

\usepackage[]{amssymb}
\usepackage{graphicx}

\begin{document}
\title{Generalized Minkowski space with changing shape.}
\author{\'A.G.Horv\'ath}
\date{2012 May}

\maketitle

\begin{abstract}
In earlier papers we changed the concept of the inner product to a more general one, to the so-called Minkowski product. This product changes on the tangent space hence we could investigate a more general structure than a Riemannian manifold. Particularly interesting such a model when the negative direct component has dimension one and the model shows certain space-time character. We will discuss this case here. We give a deterministic and a non-deterministic (random) variant of a such a model. As we showed, the deterministic model can be defined also with a ``shape function".
\end{abstract}

{\bf Keywords: generalized space-time model, normed space,  \\ Minkowski space, random and stochastic models }

\section{Introduction}

In this paper we construct a model which based on the well-known concept of Minkowski space. We generalize it in two manners. First, we change the concept of inner product to a more general concept of product which we call Minkowski product. This product changes on the tangent space hence we could investigate a more general structure than a Riemannian manifold. Particularly interesting such a model when the negative direct component has dimension one and the model shows certain space-time character. We will discuss this case here. Secondly we give a non-deterministic (random) variation of our model. We prove that in a finite range of time the random model can be approximated by a deterministic model. Thus, in calculations the deterministic model has an important role. More conveniently, it can be defined by the concept of a ``shape function". As an example it can be shown the validity of the equalities of special relativity theory. We shall public it in a forthcoming paper since the scope of the present one is still too high.

This paper is based on three previous papers of the author (\cite{gho1}, \cite{gho2}, \cite{gho3}). These contain some definitions and theorems which will be generalized here and some others which we mention and use now. We now give a short summary  for better understandability.

\subsection{Generalized space-time model}

The following definition was introduced in \cite{gho1} to give a common root of the theories of s.i.p. (semi-inner product) and i.i.p. (indefinite-inner product) spaces, respectively.

\begin{defi}
The semi-indefinite inner product (s.i.i.p.) on a complex vector space $V$ is a complex function $[x,y]:V\times V\longrightarrow \mathbb{C}$ with the following properties:
\begin{description}

\item[1] $[x+y,z]=[x,z]+[y,z]$ (additivity in the first variable),
\item[2] $[\lambda x,y]=\lambda[x,y]$ \mbox{ for every } $\lambda \in \mathbb{C}$ (homogeneity in the first variable),
\item[3] $[x,\lambda y]=\overline{\lambda}[x,y]$ \mbox{ for every } $\lambda \in \mathbb{C}$ (homogeneity in the second variable),
\item[4] $[x,x]\in \mathbb{R}$ \mbox{ for every } $x\in V$ (the corresponding quadratic form is real-valued),
\item[5] if either $[x,y]=0$ \mbox{ for every } $y\in V$ or $[y,x]=0$ for all $y\in V$, then $x=0$ (nondegeneracy),
\item[6] $|[x,y]|^2\leq [x,x][y,y]$ holds on non-positive and non-negative subspaces of V, respectively. (the Cauchy-Schwartz inequality is valid on positive and negative subspaces, respectively).
\end{description}
A vector space $V$ with a s.i.i.p. is called an s.i.i.p. space.
\end{defi}
The interest in s.i.i.p. spaces depends largely on the example spaces given by the s.i.i.p. space structure.

\noindent {\bf Example 1:} We conclude that an s.i.i.p. space is a
homogeneous s.i.p. space if and only if the property of positivity (cf. {\bf 4} in Definition 1) also holds for the product. An s.i.i.p. space is an i.i.p. space if and only if the
s.i.i.p. is an antisymmetric product. It is clear that
both of the classical "Minkowski spaces" can be represented either by an
s.i.p or by an i.i.p., so automatically they can also be represented as an
s.i.i.p. space.
Two given s.i.p. spaces can be combined into a third one. More precisely, the following statement can be proved:
\begin{lemma}[\cite{gho1}]
Let  $(S,[\cdot,\cdot]_S)$ and $(T,-[\cdot,\cdot]_T)$ be two s.i.p.
spaces. Then the function
$[\cdot,\cdot]^-:(S+T)\times(S+T)\longrightarrow \mathbb{C}$ defined
by
$$
[s_1+t_1,s_2+t_2]^-:=[s_1,s_2]-[t_1,t_2]
$$
is an s.i.p. on the vector space $S+T$.
\end{lemma}

We can define also an another product yielding a more interesting structure on $V$.

\begin{defi}[\cite{gho1}]
Let $(V,[\cdot,\cdot])$ be an s.i.i.p. space. Let $S,T\leq V$ be positive and negative subspaces, where $T$ is a direct complement of $S$ with
respect to $V$. Define a product on $V$ by the equality $[u,v]^+=[s_1+t_1,s_2+t_2]^+=[s_1,s_2]+[t_1,t_2]$, where $s_i\in S$ and $t_i\in T$,
respectively.  Then we say that the pair $(V,[\cdot,\cdot]^+)$ is a \emph{ generalized Minkowski space} with \emph{Minkowski product} $[\cdot,\cdot]^+$. We also say
that $V$ is a real generalized Minkowski space if it is a real vector space and the s.i.i.p. is a real valued function.
\end{defi}

\begin{remark}

\begin{enumerate}

\item The Minkowski product defined by the above equality satisfies properties {\bf 1}-{\bf 5} of the s.i.i.p.. But in general, property {\bf 6}
does not hold. For this, see the corresponding example in \cite{gho1}.

\item By Lemma 1 the s.i.p. $\sqrt{[v,v]^-}$ is a norm function on $V$ which can give an embedding space for a generalized Minkowski
space. This situation is analogous to the situation when a pseudo-Euclidean space is obtained from a Euclidean space by the action of an i.i.p.
\end{enumerate}
\end{remark}

\begin{defi}[\cite{gho1}]
Let $V$ be a generalized Minkowski space. Then we call a vector \emph{space-like}, \emph{light-like}, or \emph{time-like} if its scalar square is positive, zero, or negative, respectively. Let $\mathcal{S}, \mathcal{L}$ and $\mathcal{T}$ denote the sets of the space-like, light-like, and time-like vectors,
respectively.
\end{defi}

In a finite dimensional, real generalized Minkowski space with $\dim T=1$ we can geometrically characterize these sets of vectors. Such a space
is called \emph{generalized space-time model}. In this case $\mathcal{T}$ is a union of its two parts, namely
$$
 \mathcal{T}=\mathcal{T}^+\cup \mathcal{T}^-,
$$
where with respect a basis with time-like vector $e_n\in T$
$$
\mathcal{T}^+=\{s+t\in \mathcal{T} | \mbox{ where } t=\lambda e_n \mbox{ for } \lambda \geq 0\} \mbox{ and }
$$
$$\mathcal{T}^-=\{s+t\in \mathcal{T} |
\mbox{ where } t=\lambda e_n \mbox{ for } \lambda \leq 0\}.
$$
It can be proved that $\mathcal{T}$ is an open double cone with boundary $\mathcal{L}$, and the positive part $\mathcal{T}^+$ (resp. negative part $\mathcal{T}^-$) of $\mathcal{T}$ is convex.

\subsection{Convexity, fundamental forms}

Let $S$ be a continuously differentiable s.i.p. space (see in \cite{gho1}),  $V$ be a generalized space-time model and $F$ a hypersurface. We shall say that $F$ is a \emph{space-like hypersurface} if the Minkowski product is positive on its all tangent hyperplanes. The objects of this section are the convexity, the fundamental forms, the concepts of curvature, the arc-length and the geodesics. We define these concepts with respect to a generalized space-time model. With respect to a pseudo-Euclidean or a semi-Riemann space these definitions can be found e.g. in the notes \cite{moussong} and the book \cite{dubrovin}, respectively.

\begin{defi}[\cite{moussong}]
We say that a hypersurface is \emph{convex} if it lies on one side of its each tangent hyperplanes. It is \emph{strictly convex} if it is convex and its each tangent hyperplane contains precisely one point of the hypersurface.
\end{defi}

In a Euclidean space the first fundamental form is a positive definite quad\-ra\-tic form induced by the inner product of the tangent space.

In our generalized space-time model the first fundamental form is defined by the scalar square of the tangent vectors with respect to the Minkowski product restricted to the tangent hyperplane.

Let $F$ be a hypersurface defined by the function $f:S\longrightarrow V$. Here
$$
f(s)=s+\mathfrak{f}(s)e_n
$$
denotes a point of $F$. The curve $c:\mathbb{R}\longrightarrow S$ define  a curve on $F$. We assume that $c$ is a $C^2$-curve.
\begin{defi}[\cite{gho2}]
The \emph{first fundamental form} at the point $(f(c(t))$ of the hypersurface $F$  is the product
$$
\mathrm I_{f(c(t))}:=[D(f\circ c)(t),D(f\circ c)(t)]^+.
$$
\end{defi}

The variable of it is a tangent vector of a variable curve $c$ lying on $F$ through the point $(f(c(t))$.
We can see that the first fundamental form is homogeneous of the second order but (in general) it has no bilinear representation.

We introduce the unit normal vector field $n^0$ as
$$
n^0(c(t)):=\left\{\begin{array}{cc}n(c(t)) & \mbox{ if } n \mbox{ light-like vector}\\
\frac{n(c(t))} {\sqrt{|[n(c(t)),n(c(t))]^+|}} & \mbox{ otherwise. }\end{array}\right.
$$

\begin{defi}[\cite{gho2}]
The \emph{second fundamental form} at the point $f(c(t))$ is defined by one of the equivalent formulas:
$$
\mathrm I \mathrm I:=[D^2(f\circ c)(t),(n^0\circ c)(t)]^+_{(f\circ c)(t)}=-{[D(f\circ c)(t),\cdot]^+}'_{D(n^0\circ c)(t)}((n^0\circ c)(t)).
$$
\end{defi}

If we consider a $2$-plane in the tangent hyperplane then it has a two dimensional inverse image in $S$ by the regular linear mapping $Df$. The plane we get is a normed plane in which we can consider an Auerbach basis $\{e_1,e_2\}$.
\begin{defi}[\cite{gho2}]
The \emph{sectional principal curvature} of a $2$-section of the tangent hyperplane in the direction of the $2$-plane spanned by
$$
\{u=Df(e_1) \mbox{ and } v=Df(e_2)\}
$$
are the extremal values of the function
$$
\rho (D(f\circ c)):=\frac{\mathrm I \mathrm I_{f\circ c(t)}}{\mathrm I_{f\circ c(t)}},
$$
of the variable $D(f\circ c)$. We denote by $\rho(u,v)_{\mathrm{max}}$ and $\rho(u,v)_{\mathrm{min}}$ these quantities. The \emph{sectional (Gauss) curvature} $\kappa(u,v)$  (at the examined point $c(t)$) is the product
$$
\kappa(u,v):=[n^0(c(t)),n^0(c(t))]^+ \rho (u,v)_{\mathrm{max}}\rho(u,v)_{\mathrm{min}}.
$$
\end{defi}
In the case of a symmetric and bilinear product, both of the fundamental forms are quadratic ones and the sectional principal curvatures are attained in orthogonal directions.

Ricci and scalar curvatures can be defined as well.

\begin{defi}[\cite{gho2}]
The \emph{Ricci curvature} $\mathrm{Ric}(v)$ in the direction of the tangent vector $v$ at the point $f(c(t))$ is
$$
\mathrm{Ric}(v)_{f(c(t))} := (n - 2)\cdot E(\kappa_{f(c(t))}(u,v))
$$
where $\kappa_{f(c(t))}(u,v)$ is the random variable of the sectional curvatures of the two planes spanned by $v$ and a random  $u$ of the tangent hyperplane holding the equality $[u,v]^+=0$. We also say that the scalar curvature of the hypersurface $f$ at its point $f(c(t))$ is
$$
\Gamma_{f(c(t))}:= {n-1 \choose 2} \cdot E(\kappa_{f(c(t))}(u,v)).
$$
\end{defi}

The following special cases are important.

\subsubsection{Imaginary unit sphere}

The set $H:=\{ v\in V | [v,v]^+=-1\}$ is called the \emph{ imaginary unit sphere } of the generalized space-time model. $H^+$ is the connected part of $H$ defined by the function
$$
\mathfrak{h}:s\longmapsto \sqrt{1+[s,s]}.
$$

The geometric properties of $H^+$, using the differential geometry of a generalized space-time model, can be listed as follows:
\begin{itemize}
\item Let $S$ be a continuously differentiable s.i.p. space, then $(H^+,ds^2)$ is a Minkowski-Finsler space (see this concept in \cite{gho1}).
\item $H^+$ is always convex. It is strictly convex if and only if the s.i.p. space $S$ is a strictly convex space.(\cite{gho2})
\item If $S$ is a continuously differentiable s.i.p. space then $H^+$ has constant negative curvature.(\cite{gho2})
\end{itemize}

We can regard $H^+$ as a natural generalization of the usual hyperbolic space. Thus we can say that $H$ is a premanifold with constant negative curvature and $H^+$ is a \emph{prehyperbolic} space.

\subsubsection{de Sitter sphere}

The set $G$ is the collection of those points of a generalized space-time model which has scalar square equal to one. In a pseudo-Euclidean space this set was called the de Sitter sphere. The tangent hyperplanes of the de Sitter sphere are pseudo-Euclidean spaces. $G$ is not a hypersurface but we can restrict our investigation to the positive part of $G$ defined by
$$
G^+=\{s+t\in G \mbox{ : } t=\lambda e_n \mbox{ where } \lambda > 0\}.
$$
We remark that the local geometries of $G^+$ and $G$ topologically identical. $G^+$ is a hypersurface defined by the function
$$
g(s)=s+\mathfrak{g}(s)e_n,
$$
where
$$
\mathfrak{g}(s)=\sqrt{-1+[s,s]} \mbox{ for } [s,s]>1.
$$

The results on $G^+$ are the following:
\begin{itemize}
\item $G^+$ and its tangent hyperplanes are intersecting, consequently there is no point at which $G$ would be convex.(\cite{gho2})
\item The de Sitter sphere $G$ has constant positive curvature if $S$ is a continuously differentiable s.i.p space.(\cite{gho2})
\end{itemize}

On the basis of this theorem we can tell about $G$ as a premanifold of constant positive curvature and we may say that it is a \emph{presphere}.

\subsubsection{The light cone}

The inner geometry of the light cone $L$ can be determined, too. Let $L^+$ be the positive part of the double cone determined by the function:
$$
l(s)=s+\sqrt{[s,s]}e_n.
$$

\begin{itemize}
\item The light cone $L^+$ has zero curvatures if $S$ is a continuously differentiable s.i.p space.(\cite{gho2})
\end{itemize}

Hence $L$ is a premanifold with zero sectional, Ricci and scalar curvatures, respectively. We may also say that it is a \emph{pre-Euclidean} space.

\subsubsection{The unit sphere of the s.i.p. space $(V,[\cdot,\cdot]^-)$}

The set $K$ collects the points of the unit sphere of the embedding s.i.p. space. In a pseudo-Euclidean space it is the unit sphere of the embedding Euclidean space. Its tangent hyperplanes are pseudo-Euclidean spaces. $K$ is not a hypersurface but we can also restrict our investigation to its positive part defined by
$$
K^+=\{s+t\in K \mbox{ : } t=\lambda e_n \mbox{ where } \lambda > 0\}.
$$
Hence it can be defined by the function:
$$
k(s)=s+\mathfrak{k}(s)e_n,
$$
where
$$
\mathfrak{k}(s)=\sqrt{1-[s,s]} \mbox{ for } [s,s]<1.
$$

The basic properties of $K^+$ are
\begin{itemize}
\item $K^+$ is convex. If $S$ is a strictly convex space, then $K^+$ is also strictly convex.
\item The fundamental forms are
$$
\mathrm I =[\dot{c},\dot{c}]-\frac{\left([\dot{c}(t),c(t)]+ [c(t),\cdot]'_{\dot{c}(t)}(c(t))\right)^2}{4(1-[c(t),c(t)])}= [\dot{c},\dot{c}]-\frac{[\dot{c}(t),c(t)]^2}{1-[c(t),c(t)]},
$$
$$
\mathrm I \mathrm I=\frac{1}{\sqrt{|-1+2[c(t),c(t)]|}}\left( -[\dot{c}(t),\dot{c}(t)]+\frac{[\dot{c}(t),c(t)]^2}{-1+[c(t),c(t)]}\right)=
$$
$$
= -\frac{1}{\sqrt{|-1+2[c(t),c(t)]|}}\mathrm I .
$$
\item
The principal, sectional, Ricci and scalar curvatures at a point $k(c(t))$ are
$$
\rho_{\max}(u,v)=\rho_{\min}(u,v)=-\frac{1}{\sqrt{|-1+2[c(t),c(t)]|}},
$$
$$
\kappa(u,v):=[n^0(c(t)),n^0(c(t))]^+ \rho (u,v)_{\mathrm{max}}\rho(u,v)_{\mathrm{min}}=
\frac{1}{-1+2[c(t),c(t)]},
$$
$$
\mathrm{Ric}(v)_{k(c(t))} := (n - 2)\cdot E(\kappa_{k(c(t))}(u,v))=\frac{n-2}{-1+2[c(t),c(t)]},
$$
and
$$
\Gamma_{k(c(t))}:= {n-1 \choose 2} \cdot E(\kappa_{f(c(t))}(u,v))=
\frac{{n-1 \choose 2}}{-1+2[c(t),c(t)]},
$$
respectively.
\item
Finally we remark that at the points of $K^+$ having the equality
$$
2[c(t),c(t)]=1
$$
all of the curvatures can be defined as in the case of the light cone and can be regarded as zero.
\end{itemize}

\section{The absolute time}

We assume that there is an absolute coordinate system of dimension $n$ in which we are modeling the universe by a time-space model. The origin is a generalized space-time model in which the time axis plays the role of the absolute time.  Its points are unattainable and immeasurable for me and the corresponding line is also in the exterior of the modeled universe. We note that in the Minkowskian space-time it was assumed only on the axes determining the space-coordinates. This means that in our model, even though the axis of time belongs to the double cone of time-like points, its points do not belong to the modeled universe. In a fixed moment (with respect to this absolute time) the collection of the points of the space can be regarded as an open ball of the embedding normed space centered at the origin that does not contain the origin. The omitted point is the origin of a coordinate system giving the space-like coordinates of the world-points with respect to our time-space system. Since the points of the axis of the absolute-time are not in our universe there is no reference system in our modeled world which determines the absolute time.

First we need a probability measure which describes the change of the shape of the model. We regard this change random in the absolute time and as a perturbation of normed spaces which are ``almost Euclidean space".

\subsection{The probability space of norms}

The distance of two normed spaces can be measured by the Hausdorff distance of their unit balls. This motivated the investigations of \cite{gho3}. We recall it in this section. Every norm function of a real, finite-dimensional normed space $V$ can be defined by its unit ball. In a Cartesian coordinate system of the Euclidean vector space $(V, \langle \cdot, \cdot \rangle)$ with origin $O$ it is convex body centrally symmetric about $O$ (shortly $O$-symmetric). Such bodies form a closed proper subset $\mathcal{K}_0$ of the space of convex bodies of the Euclidean vector space. It is known that Hausdorff distance (denoted by $\delta_H$) is a metric on this space and with this metric the space $(\mathcal{K},\delta_{h})$ is a locally compact one (see \cite{gruber 1},\cite{gruber 2}). In \cite{bandt} it was proved that there is no positive $\sigma$-finite Borel measure on it which is invariant with respect to all isometries of $(\mathcal{K},\delta_h)$ into itself. In paper \cite{hoffmann} was proved that each $\sigma$-finite rotation and translation
invariant Borel measure on $(\mathcal{K},\delta_{h})$ is the vague limit of such measures and that each $\sigma $-finite
Borel measure on $(\mathcal{K},\delta_{h})$ is the vague limit of measures of the form
$$
\sum\limits_{i=1}^{\infty} \alpha_n\delta_{K_n},
$$
where $\{K_n \mbox{ , } n\in \mathbb{N}\}$ is a countable, dense subset of $(\mathcal{K},\delta_{h})$, $(\alpha_n)$ is a sequence of positive real numbers for which $\sum \limits_{i=1}^\infty \alpha_n <\infty $ and $\delta_{K_n}$ denote the Dirac measure concentrated at $K_n$.

In \cite{gho3} it was proved that on the space of centrally symmetric convex bodies there can be given such a geometric probability measure $P$ which has additionally the following property: its pushforward measure $\alpha_0(K)^{-1}(P)$ by the thinness mapping
$$
\alpha_0(K)=\frac{d(K)}{w(K)+d(K)}
$$
has truncated normal distribution on the interval $[\frac{1}{2},1)$. (Here $w(K)$ and $d(K)$ are the width and the diameter of the body $K$, respectively.) This measure was constructed step by step in the following manner:

Let $B_E$ be the unit ball of the embedding Euclidean space and let
$$
\mathcal{K}_{0}^1:=\{ K\in \mathcal{K}_{0} \mid \delta^h(K,B_E)=1\}
$$
be the unit sphere around $B_E$ with respect to the Hausdorff metric.
The space $\widetilde{\mathcal{K}_0^1}$ collects the representatives of the classes of congruent bodies of the space $\mathcal{K}_0^1$.

First, we constructed a measure on the space $\widetilde{\mathcal{K}_0^1}$ such that its pushforward measure by the function $w$ has uniform distribution. Than proved that the direct product of this measure with the Haar measure of the group of orthogonal transformations has  analogous property; its pushforward measure by $\alpha_0(K)$ uniformly distributes on its range interval. Finally, considering an arbitrary probability measure on $[0,\infty)$ and multiplying it with the one constructed above, we get a probability measure on $\mathcal{K}_0$. From this measure by a suitable density function we can obtain a new probability measure such that its pushforward measure by the function $\alpha_0$ has truncated normal distribution (see Theorem 2, Theorem 3 in \cite{gho3}).

\begin{defi}
We say that the a probability measure is \emph{a geometric measure with normal pushforward} if the following properties hold
\begin{itemize}
\item it is invariant under orthogonal transformations of the space of norms;
\item the set of polytopes, the set of smooth bodies and a neighborhood have zero measure, positive measure and positive measure, respectively;
\item there is a natural function on the space of norms to an interval of the real line for which the pushforward of the measure has truncated normal distribution of its range interval. (Of course here we assume that the mean of the pushforward distribution is attained at the image of the unit ball of the Euclidean space.)
\end{itemize}

\end{defi}
In this paper we use always geometric measure with normal pushforward.

\subsection{Deterministic and random time-space models}

In our probabilistic model (based on a generalized space-time model) the absolute coordinates of points are calculated by a fixed basis of the embedding vector space. The vector $s(\tau)$ means the collection of the space-components with respect to the absolute time $\tau$, the quantity $\tau$ has to be measured on a line $T$ which orthogonal to the linear subspace $S$ of the vectors $s(\tau)$. (The orthogonality was considered as the Pythagorean orthogonality of the embedding normed space.) Consider a fixed Euclidean vector space with unit ball $B_E$ on $S$ and use its usual functions e.g.  volume, diameter, width, thinness and Hausdorff distance. With respect to the moment $\tau$ of the absolute time we have a unit ball $K(\tau)$ in the corresponding normed space  $\{S,\|\cdot\|^{\tau}\}$. The modeled universe at $\tau$ is the ball $\tau K(\tau)\subset \{S,\|\cdot\|^{\tau}\}$.
The shape of the model at the moment $\tau$ depends on the shape of the centrally symmetric convex body $K(\tau)$. The center of the model is on the axis of the absolute time, it cannot be determined. For calculations on time-space we need further smoothness properties on $K(\tau)$. These are
\begin{itemize}
\item $K(\tau)$ is a centrally symmetric, convex, compact, $C^2$ body of volume $\mathrm{vol}(B_E)$.
\item For each pairs of points $s',s''$ the function
$$
K:\mathbb{R}^+\cup \{0\}\rightarrow \mathcal{K}_0 \mbox{ , }\tau\mapsto K(\tau)
$$
holds the property that $[s',s'']^{\tau}:\tau\mapsto [s',s'']^{\tau}$ is a $C^1$-function.
\end{itemize}

\begin{defi}
We say that a generalized space-time model endowed with a function $K(\tau)$ holding the above properties is a \emph{deterministic time-space model}.
\end{defi}

The main subset of a deterministic time-space model contains the points of negative norm-square. This is the set of time-like points and the upper connected sheet of the time-like points is the modeled universe. The points of the universe have positive time-components. We denote this model by
$
\left(M,K(\tau)\right).
$

We remark that in the two-dimensional case for each $\tau$, $K(\tau)$ is a segment with length two, thus our model is the $2$-dimensional space-time. On the other hand, with $n$ greater than or equal to $3$, the two-dimensional space-time sections of our general space-time bounded by general (non-convex) curves symmetric about the time-axis (see on Fig.1).

\begin{figure}[htbp]
\centerline{\includegraphics[scale=1]{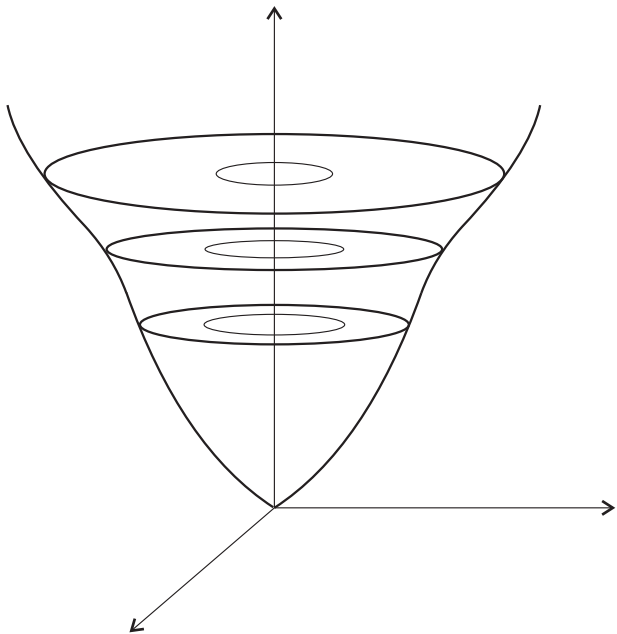}}
\caption{The shape of the universe.}
\end{figure}

Of course, we should choose the function $K(\tau)$ ``randomly". To this purpose we use Kolmogorov's extension theorem (or theorem on consistency, see in \cite{kolmogorov}). This says that a suitably "consistent" collection of finite-dimensional distributions will define a probability measure on the product space. The sample space here is $\mathcal{K}_0$ with the Hausdorff distance. It is a locally compact, separable (second-countable) metric space. By Blaschke's selection theorem (see in \cite{gruber 1}) $\mathcal{K}$ is a boundedly compact space so it is also complete. It is easy to check that $\mathcal{K}_0$ is also a complete metric space if we assume that the non-proper bodies (centrally symmetric convex compact sets with empty interior) also belong to it. In the remaining part we regard such a body as the unit ball of a normed space of smaller dimension. Finally, let $P$ be the probability measure defined in Subsection 3.1. In every moment we consider the same probability space $\left(\mathcal{K}_0, P\right)$ and also consider in each of the finite collections of moments the corresponding product spaces $\left((\mathcal{K}_0)^r, P^r\right)$ . The consistency assumption of Kolmogorov's theorem now automatically holds. By the extension theorem we have a probability measure $\hat{P}$ on the measure space of the functions on $T$ to $\mathcal{K}_0$ with the $\sigma$-algebra generated by the cylinder sets of the space. The distribution of the projection of $\hat{P}$ to the probability space of a fix moment is the distribution of $P$.

\begin{defi}
Let $(K_\tau \mbox{ , }\tau\geq 0)$ be a random function defined as an element of the Kolmogorov's extension $\left(\Pi \mathcal{K}_0, \hat{P}\right)$ of the probability space $\left(\mathcal{K}_0, P\right)$. We say that the generalized space-time model with the random function
$$
\hat{K}_\tau:=\sqrt[n]{\frac{\mathrm{ vol}(B_E)}{\mathrm{ vol}(K_\tau)}}K_\tau
$$
is a \emph{random time-space model}. Here $\alpha_0(K_\tau)$ is a random variable with truncated normal distribution and thus $(\alpha_0(K_\tau) \mbox{ , } \tau\geq 0)$ is a stationary Gaussian process. We call it the \emph{shape process} of the random time-space model.
\end{defi}
It is clear that a deterministic time-space model is a special trajectory of the random time-space model. The following theorem is essential.
\begin{theorem}
For a trajectory $L(\tau)$ of the random time-space model, for a finite set $0\leq \tau_1\leq \cdots \leq \tau_s$  of  moments and for a $\varepsilon >0$ there is a deterministic time-space model defined by the function $K(\tau)$ for which
$$
\sup\limits_{i}\{\rho_H\left(L(\tau_i), K(\tau_i)\right)\} \leq \varepsilon.
$$
\end{theorem}

\proof Since the set of centrally symmetric convex bodies with $C^\infty$-boundary is dense in the set of centrally symmetric convex bodies (see \cite{schneider}), we can choose, for every $\tau_i$, a body $K(\tau_i)\in \mathcal{K}_0$ with $C^2$ boundary with the required volume for which
$$
\rho_H\left(L(\tau_i), K(\tau_i)\right) \leq \varepsilon
$$
holds. We shall prove that these bodies can be connected with such a trajectory of the random time-space model for which the function $K$ holds the properties of the defining function of a deterministic time-space model. The impact of the $K$ function on a fix vector $s\in S$ can be checked on the vary of its norm. Using the Minkowski functional, we can get the norm of a vector $s$ as the length of a fixed segment relative to the length of the diameter of the unit ball intersected by the half-line containing the segment $[O,P]$. (Here $O$ and $P$ are the origin and the endpoint of the vector $s$, respectively.) This means that we can determine of the vary of the length of a diameter of a fixed direction if we vary the shape of the body by the time. Consider a representation of the body by polar coordinates with respect to its center $O$. Since the boundary of the body is of class $C^2$, all of their coordinate functions have the analogous property. This function depends also on the time $\tau$, the change of the unit ball implies the change of its coordinate functions. We say that the trajectory $K(\tau)$ is a continuously differentiable function if for a fixed coordinate representation its coordinate functions are continuously differentiable functions of the time. This is equivalent to the property that the support function $h_{(K(\tau))}$  is continuously differentiable as the function of the time $\tau$. The differentiability property of the trajectory implies the analogous differentiability property of the change of the norm of a fix vector since the points of the boundary of the unit ball has an equation of the form
$$
r^\tau=\left(r(\varphi_1,\cdots ,\varphi_{n-1})\right)^{\tau}.
$$
We can conclude that if the trajectory $K(\tau)$ is a continuously differentiable function, this holds also for the function
$$
\tau\rightarrow \sqrt{[s,s]^\tau}.
$$
In a space $S$ with an inner product the polarity equation implies the required assumption. If $S$ is (only) a smooth normed space with a semi inner product, we need further comments.
Since for a differentiable norm function McShane's equality holds (see \cite{gho2}), we have
$$
[x,y]^\tau= \|y\|^\tau (\left(\|\cdot\|^\tau\right)'_x(y))= \|y\|^\tau (\|\cdot\|'_x(y))^\tau.
$$
On the other hand, the function $(\|\cdot\|'_x(y))^\tau$ is also continuously differentiable function of $y$, thus the thread using on the norm function above is applicable for it, too. This means that the differentiability property of the trajectory implies the analogous differentiability property of the function
$$
\tau\rightarrow (\|\cdot\|'_x(y))^\tau.
$$
Using the rule of the product function we also have that $\tau\rightarrow [x,y]^\tau$ is continuously differentiable if the trajectory
$$
\tau \rightarrow K(\tau)
$$
holds this property.

We now define a differentiable trajectory through the points $\left(\tau_i,K(\tau_i)\right)$. If  $\tau,\tau'_i\in[\tau_i,\tau_{i+1}]$ denote by $K_{Bezier}(\tau)$ the formal Bezier spline of second order through the points $\left(\tau_i,K(\tau_i)\right)$ and $\left(\tau_{i+1},K(\tau_{i+1})\right)$ with "tangents" through the point $\left(\tau'_i,L(\tau'_i)\right)$. Thus we have by definition
$$
K_{Bezier}(\tau):=\left(1 - \frac{\tau-\tau_i}{\tau_{i+1}-\tau_{i}}\right)^{2}K(\tau_i) +
$$
$$
+2\left(1 - \frac{\tau-\tau_i}{\tau_{i+1}-\tau_{i}}\right)\frac{\tau-\tau_i}{\tau_{i+1}-\tau_{i}}L(\tau_i)+ \left(\frac{\tau-\tau_i}{\tau_{i+1}-\tau_{i}}\right)^{2}K(\tau_{i+1}),
$$
where the addition is the Minkowski addition and the product is the respective homothetic mapping.
If we assume that for all values of $i$ ($1<i<s$) the body $K(\tau_i)$ is a Minkowski convex combination of the bodies $L(\tau'_i)$ and $L(\tau'_{i+1})$ the function $K_{Bezier}(\tau)$ is valid on the whole interval $[\tau_1,\tau_{s}]$. Since for positive constants $\alpha$, $\beta$ we have
$$
h_{\alpha K'+\beta K''}(x)=\alpha h_{K' }(x)+\beta h_{K''}(x),
$$
we also get that $K_{Bezier}(\tau)$ is a continuously differentiable trajectory in its whole domain. We have to prove yet that for a fixed $\tau$, the set $K_{Bezier}(\tau)$ is a centrally symmetric convex compact body with $C^2$-class boundary but these statements follow immediately from the concept of Minkowski linear combination.

Finally we normalize this trajectory under the volume function and extract it to the whole $T$. The function $K(\tau)$ determines a required deterministic time-space model if we define it as follows:
$$
K(\tau)=\left\{
\begin{array}{lcl}
\sqrt[n]{\frac{\mathrm{ vol}(B_E)}{\mathrm{ vol}(K_{Bezier}(\tau_s))}}K_{Bezier}(\tau_s)& \mbox{ if } & \tau_s<\tau\\
\sqrt[n]{\frac{\mathrm{ vol}(B_E)}{\mathrm{ vol}(K_{Bezier}(\tau))}}K_{Bezier}(\tau)& \mbox{ if } & \tau_1\leq \tau \leq \tau_s \\
\sqrt[n]{\frac{\mathrm{ vol}(B_E)}{\mathrm{ vol}(K_{Bezier}(\tau_1))}}K_{Bezier}(\tau_1)& \mbox{ if } & \tau<\tau_1 \mbox{ . }
\end{array}\right.
$$
\qed

An important consequence of Theorem 1 is then that without loss of generality we can assume, that the time-space model is deterministic.

\subsection{Product in a deterministic time-space model}

We can give a product similar to the Minkowski product of a generalized space-time model. In a two-dimensional plane the role of the light-cone play the curve
$$
\left[\alpha^e(\tau) e,\alpha^e(\tau) e\right]^{\tau}+\left[\tau,\tau\right]=0.
$$
For a fixed direction $x$, we consider the curves
$$
t_{\beta,e}:\tau\mapsto \beta\alpha^e(\tau) e+\tau e_n
$$
through the point $x=\beta\alpha^e(\tau) e+\tau e_n$. Note that
$x$ is a time-like point if $|\beta|<1$. The role of the imaginary unit sphere is played by the set of points
$$
\cup\left\{\left\{s+\tau \mbox{ where } \sqrt{[s,s]^\tau+1}=\tau \right\} \mbox{ , } \tau\geq 1\right\}.
$$
In the direction of $e$ it is a curve defined by the implicit equation
$$
\sqrt{[s,s]^\tau+1}=\tau \mbox{ , } \tau\geq 1
$$
The intersection of this curve with $t_{\beta,e}$ is a point satisfying the equality
$$
[\beta\alpha^e(\tau ^\star) e,\beta\alpha^e(\tau^\star) e]^{\tau^\star}+1=\left(\tau^\star\right)^2,
$$
with parameter $\tau^\star$,
and hence we get
$$
\beta^2 \left(\tau^\star\right)^2+1=\left(\tau^\star\right)^2,
$$
or equivalently
$$
\left(\tau^\star\right)^2=\frac{1}{1-\beta^2}.
$$
Assuming that our examination is on the positive part of the set of time-like points we have
$$
\tau^\star=\frac{1}{\sqrt{1-\beta^2}} \mbox{ or } \beta=\frac{\sqrt{(\tau^\star)^2-1}}{\tau^\star}.
$$
In the space-time model the tangent of the imaginary unit curve is orthogonal to the position vector of the common point. This requires that in the case of generalized space-time model, the product
$$
\left[e+\left(\sqrt{[s,s]^\tau+1}\right)'_e\left(\beta\alpha^e(\tau^\star)e\right)e_n,\beta\alpha^e\left(\tau^\star\right)e+\tau^\star e_n\right]
$$
will be equal to zero.
Another claim that the product is equal to the corresponding norm-square in the case when its arguments contains the same vectors. We will need a lemma on the directional derivative of the function which defines the imaginary unit sphere.

\begin{lemma}
The directional derivative of the real valued function
$$
\mathfrak{h}(s)=\sqrt{[s,s]^{\mathfrak{h}(s)}+1}
$$
is
$$
\mathfrak{h}'_{e}(s)=\left(1- \frac{\frac{\partial{\left[s,s \right]^{\tau}}}{\partial \tau}(\mathfrak{h}(s))}{2\sqrt{1+[s,s]^{\mathfrak{h}(s)}}}\right)^{-1}\frac{[e,s]^{\mathfrak{h}(s)}}{\sqrt{1+[s,s]^{\mathfrak{h}(s)}}}=
$$
$$
=\frac{2}{2\mathfrak{h}(s)-\frac{\partial{\left[s,s \right]^{\tau}}}{\partial \tau}(\mathfrak{h}(s))}[e,s]^{\mathfrak{h}(s)},
$$
or equivalently
$$
\mathfrak{h}'_{e}(s)=\frac{1}{\mathfrak{h}(s)-\|s\|^{\mathfrak{h}(s)}\frac{\partial{\|s\|^{\tau}}}{\partial \tau}(\mathfrak{h}(s))}[e,s]^{\mathfrak{h}(s)}.
$$

\end{lemma}
\proof
The considered derivative is
$$
\mathfrak{h}'_e(s)=\frac{1}{2\sqrt{1+[s,s]^{\mathfrak{h}(s)}}}([s,s]^{\mathfrak{h}(s)})'_e.
$$
It can be seen easily (or use the calculation of Theorem 1 with the substitutions $c(t+\lambda)=s+\lambda e$, $(f_1)_S=(f_2)_S=\mathrm{id}|_S$ and $(f_1)_T=(f_2)_T=\mathfrak{h}$) that the directional derivative in question (independently of the sought product) is equal to
$$
\frac{1}{2\sqrt{1+[s,s]^{\mathfrak{h}(s)}}}\left([e,s]^{\mathfrak{h}(s)}+\left([s,\cdot]^{\mathfrak{h}(s)}\right)'_e(s)+ \frac{\partial{\left[s,s \right]^{\tau}}}{\partial \tau}(\mathfrak{h}(s))\cdot(\mathfrak{h})'_e(s)\right)=
$$
$$
=\frac{1}{2\sqrt{1+[s,s]^{\mathfrak{h}(s)}}}\left(2[e,s]^{\mathfrak{h}(s)}+ \frac{\partial{\left[s,s \right]^{\tau}}}{\partial \tau}(\mathfrak{h}(s))\cdot(\mathfrak{h})'_e(s)\right).
$$
Thus we get
$$
\mathfrak{h}'_e(s)\left(1- \frac{\frac{\partial{\left[s,s \right]^{\tau}}}{\partial \tau}(\mathfrak{h}(s))}{2\sqrt{1+[s,s]^{\mathfrak{h}(s)}}}\right)=\frac{[e,s]^{\mathfrak{h}(s)}}{\sqrt{1+[s,s]^{\mathfrak{h}(s)}}},
$$
or equivalently the required formulas
$$
\mathfrak{h}'_e(s)=\left(1- \frac{\frac{\partial{\left[s,s \right]^{\tau}}}{\partial \tau}(\mathfrak{h}(s))}{2\sqrt{1+[s,s]^{\mathfrak{h}(s)}}}\right)^{-1}\frac{[e,s]^{\mathfrak{h}(s)}}{\sqrt{1+[s,s]^{\mathfrak{h}(s)}}}=
$$
$$
=\frac{1}{\mathfrak{h}(s)-\|s\|^{\mathfrak{h}(s)}\frac{\partial{\|s\|^{\tau}}}{\partial \tau}(\mathfrak{h}(s))}[e,s]^{\mathfrak{h}(s)}.
$$
\qed

Now $s$ and $\mathfrak{h}(s)$ are equals to $\beta\alpha^e(\tau^\star)e$ and $\tau^\star$, respectively. We get that
$$
\left(\sqrt{[s,s]^\tau+1}\right)'_e\left(\beta\alpha^e(\tau^\star)e\right)=\frac{\beta\alpha^e(\tau^\star)[e,e]^{\tau^\star}}{\tau^\star-\beta\alpha^e(\tau^\star) \|e\|^{\tau^\star}\frac{\partial{\|\beta\alpha^e(\tau^\star)e\|^{\tau}}}{\partial \tau}(\tau^\star)}=
$$
$$
=\frac{\beta\alpha^e(\tau^\star)[e,e]^{\tau^\star}}{\tau^\star\left(1-\beta\frac{\partial{\|\beta\alpha^e(\tau^\star)e\|^{\tau}}}{\partial \tau}(\tau^\star)\right)}.
$$
Any natural concept of product should satisfy the basic property of the Min\-kows\-ki product. Thus we assume that the unknown product the following equality holds:
$$
\left[e+\left(\sqrt{[s,s]^\tau+1}\right)'_e\left(\beta\alpha^e(\tau^\star)e\right)e_n,\beta\alpha^e\left(\tau^\star\right)e+\tau^\star e_n\right]^?=
$$
$$
=[e,\beta\alpha^e(\tau^\star)e]^{\tau^\star}-\frac{\beta\alpha^e(\tau^\star)[e,e]^{\tau^\star}}{1- \beta\frac{\partial{\|\beta\alpha^e(\tau^\star)e\|^{\tau}}} {\partial \tau}(\tau^\star)}=
$$
$$
=\frac{[e,\beta\alpha^e(\tau^\star)e]^{\tau^\star}\left(-\beta\frac{\partial{\|\beta\alpha^e(\tau^\star)e\|^{\tau}}} {\partial \tau}(\tau^\star)\right)}{1- \beta\frac{\partial{\|\beta\alpha^e(\tau^\star)e\|^{\tau}}} {\partial \tau}(\tau^\star)},
$$
showing that we lost an important orthogonality property, which was between the position and tangent vectors of the imaginary unit sphere.
On the other hand, this formula, in the case when the norm is constant, gives back this property. We have another interesting observation, which suggests that we should go on in this natural way. We try to substitute the position vector of the imaginary sphere with the tangent vector of the time axis $t_{\beta,e}$. This is the vector
$$
\tau^\star\left(\beta\frac{\partial{\alpha^e(\tau)}}{\partial\tau}(\tau^\star)e+e_n\right)=\left(\frac{\beta \tau^\star}{\sqrt{[e,e]^{\tau^\star}}}-\frac{1}{2}\frac{\beta \left(\tau^\star\right)^2\frac{\partial [e,e]^{\tau}}{\partial{\tau}}(\tau^\star)}{\sqrt{[e,e]^{\tau^\star}}[e,e]^{\tau^\star}}\right)e+\tau^\star e_n,
$$
and the product is
$$
\left[e+\left(\sqrt{[s,s]^\tau+1}\right)'_e\left(\beta\alpha^e(\tau^\star)e\right)e_n, \tau^\star\left(\beta\frac{\partial{\alpha^e(\tau)}}{\partial\tau}(\tau^\star)e+e_n\right)\right]=
$$
$$
=\beta\tau^\star\sqrt{[e,e]^{\tau^\star}}-\frac{1}{2}\frac{\beta \left(\tau^\star\right)^2\frac{\partial [e,e]^{\tau}}{\partial{\tau}}(\tau^\star)}{\sqrt{[e,e]^{\tau^\star}}}-\frac{\beta \left(\tau^\star\right)^2\sqrt{[e,e]^{\tau^\star}}}{\tau^\star -\frac{1}{2}\beta^2\left(\alpha^e(\tau^\star)\right)^2\frac{\partial [e,e]^{\tau}}{\partial{\tau}}(\tau^\star)}=
$$
$$
=\frac{-\frac{1}{2}\frac{\partial[e,\beta\alpha^e(\tau^\star)]^\tau}{\partial \tau}(\tau^\star)\left((\tau^\star)^2-\beta^2(\tau^\star)^2-\frac{1}{2}\tau^{\star}\beta^2(\alpha^e(\tau^\star))^2\frac{\partial[e,e]^\tau}{\partial \tau}(\tau^\star)\right)}{\tau^\star -\frac{1}{2}\beta^2\left(\alpha^e(\tau^\star)\right)^2\frac{\partial [e,e]^{\tau}}{\partial{\tau}}(\tau^\star)}.
$$
Using the connection among the values of $\beta $, $\tau^{\star}$ and $\alpha^e(\tau^\star)$ we get that it is zero if and only if
$$
(\tau^\star)^2-\beta^2(\tau^\star)^2=1=\frac{1}{2}\tau^{\star}\beta^2(\alpha^e(\tau^\star))^2\frac{\partial[e,e]^\tau}{\partial \tau}(\tau^\star)=\frac{1}{2}\frac{(\tau^\star)^3-\tau^\star}{[e,e]^{\tau^\star}}\frac{\partial[e,e]^\tau}{\partial \tau}(\tau^\star).
$$
This is a separable differential equation in the function $[e,e]^{\tau}$ with solution
$$
[e,e]^{\tau}=\left(1-\frac{1}{\tau^2}\right)c_e^2
$$
where $c_e$ is a constant depending on the direction $e$. This shows that by the following definition there is a non-trivial solution of the problem: Determine the time-dependence of the norm in such a way that the imaginary unit sphere intersects the time-axes $t_{\beta,e}$ orthogonally!

\begin{defi}
For two vectors $s_1+\tau_1$ and $s_2+\tau_2$
of the deterministic time-space model define their product with the equality
$$
[s_1+\tau_1,s_2+\tau_2]^{+,T}:=[s_1,s_2]^{\tau_2}+\left[\tau_1,\tau_2\right]=
$$
$$
=[s_1,s_2]^{\tau_2}-\tau_1\tau_2.
$$
\end{defi}
Here $[s_1,s_2]^{\tau_2}$ means the s.i.p defined by the norm $\|\cdot\|^{\tau_2}$. This product is not a Minkowski product, as there is no homogeneity property in the second variable.
On the other hand the additivity and homogeneity properties of the first variable, the properties on non-degeneracy of the product are again hold, and the continuity and differentiability properties of this product also remain the same as of a Minkowski product. The calculations in a generalized space-time model basically depend on a rule on the differentiability of the second variable of the Minkowski product. Introducing the notation
$$
{[f_1(c(t)),\cdot]^+}'_{D(f_2\circ c)(t)}(f_2(c(t))):=
$$
$$
:=\left([(f_1)_S(c(t)),\cdot]'_{D((f_2)_S\circ c)(t)}((f_2)_S(c(t))) - (f_1)_T(c(t))((f_2)_T\circ c)'(t)\right),
$$
we stated in \cite{gho2} (see Lemma 4), that
if $f_1, f_2: S\longrightarrow V=S+T$ are two $C^2$ maps and $c:\mathbb{R}\longrightarrow S$ is an arbitrary $C^2$ curve then
$$
([(f_1\circ c)(t)),(f_2\circ c)(t))]^+)'=
$$
$$
=[D(f_1\circ c)(t),(f_2\circ c)(t))]^++{[(f_1\circ c)(t)),\cdot]^+}'_{D(f_2\circ c)(t)}((f_2\circ c)(t)).
$$
Regarding to the importance of this rule we reproduce it in a time-space model.
Let denote by $f_S$ and $f_T$ the component functions of $f$ with respect to the subspaces $S$ and $T$, respectively. By definition, let us denote
$$
\left({[f_1(c(t)),\cdot]^{+,T}}\right)'_{D(f_2\circ c)(t)}(f_2(c(t))):=
$$
$$
=\left([(f_1)_S(c(t)),\cdot]^{(f_2)_T(c(t))}\right)'_{D((f_2)_S\circ c)(t)}((f_2)_S(c(t))) - (f_1)_T(c(t))((f_2)_T\circ c)'(t)+
$$
$$
+(f_1)_T(c(t))\frac{\partial ^2[(f_2)_S(c(t)),(f_2)_S(c(t))]^\tau}{\left(\partial \tau\right)^2}((f_2)_T(c(t)))\times 
$$
$$
\times \left[D((f_2)_S\circ c)(t),(f_2)_S(c(t))\right]^{(f_2)_T(c(t))}.
$$
We now generalize the formula of Lemma 2.
\begin{theorem}
If $f_1, f_2: S\longrightarrow V=S+T$ are two $C^2$ maps and $c:\mathbb{R}\longrightarrow S$ is an arbitrary $C^2$ curve then
$$
([(f_1\circ c)(t)),(f_2\circ c)(t))]^{+,T})'=
$$
$$
=[D(f_1\circ c)(t),f_2(c(t))]^{+,T} +\left({[f_1(c(t)),\cdot]^{+,T}}\right)'_{D(f_2\circ c)(t)}(f_2(c(t)))+
$$
$$
+\frac{\partial{\left[(f_1)_S(c(t)),(f_2)_S(c(t)) \right]^{\tau}}}{\partial \tau}((f_2)_T(c(t)))\cdot((f_2)_T\circ c)'(t)
$$

\end{theorem}

\proof
By definition
$$
([f_1\circ c,f_2\circ c)]^{+,T})'|_t:=
$$
$$
=\lim\limits_{\lambda \rightarrow 0}\frac{1}{\lambda}\left([f_1(c(t+\lambda)),f_2(c(t+\lambda ))]^{+,T}-[f_1(c(t)),f_2(c(t))]^{+,T}\right)=
$$
$$
=\lim\limits_{\lambda \rightarrow 0}\frac{1}{\lambda}\left([(f_1)_S(c(t+\lambda)),(f_2)_S(c(t+\lambda ))]^{(f_2)_T(c(t+\lambda))}-\right.
$$
$$
\left.-[(f_1)_S(c(t)),(f_2)_S(c(t))]^{(f_2)_T(c(t))}\right)+
$$
$$
+\lim\limits_{\lambda \rightarrow 0}\frac{1}{\lambda}\left([(f_1)_T(c(t+\lambda)),(f_2)_T(c(t+\lambda))] -[(f_1)_T(c(t)),(f_2)_T(c(t))]\right).
$$

The first part can be written in the form
$$
\lim\limits_{\lambda \rightarrow 0}\frac{1}{\lambda}\left([(f_1)_S(c(t+\lambda))-(f_1)_S(c(t)),(f_2)_S(c(t+\lambda ))]^{(f_2)_T(c(t+\lambda))}+\right.
$$
$$
\left.+[(f_1)_S(c(t)),(f_2)_S(c(t+\lambda))]^{(f_2)_T(c(t+\lambda))}-[(f_1)_S(c(t)),(f_2)_S(c(t))]^{(f_2)_T(c(t))}\right).
$$
We prove that it is equal to
$$
[D((f_1)_S\circ c)|_{t},(f_2)_S(c(t))]^{(f_2)_T(c(t))}+
$$
$$
+\left([(f_1)_S(c(t)),\cdot]^{(f_2)_T(c(t))}\right)'_{D((f_2)_S\circ c)(t)}((f_2)_S(c(t)))+
$$
$$
+\frac{\partial{\left[(f_1)_S(c(t)),(f_2)_S(c(t)) \right]^{\tau}}}{\partial \tau}((f_2)_T(c(t)))\cdot((f_2)_T\circ c)'(t).
$$
In this latter equation the first term comes from the value of the first bracket of the earlier one. We calculate now the remaining substraction. For this, take the fixed (absolute) coordinate system $\{e_1,\cdots ,e_{n-1}\}$ of $S$ and consider the coordinate-wise representation
$$
(f_2)_S\circ c=\sum\limits_{i=1}^{n-1}((f_2)_S\circ c)_ie_i
$$
of $(f_2)_S\circ c$. Using Taylor's theorem for the coordinate functions we have that there are real parameters $t_i \in (t,t+\lambda)$, for which
$$
((f_2)_S\circ c)(t+\lambda)=((f_2)_S\circ c)(t)+\lambda D((f_2)_S\circ c)(t)+\frac{1}{2}\lambda ^2\sum\limits_{i=1}^{n-1}((f_2)_S\circ c)''_i(t_i)e_i.
$$
Thus we get that
$$
\left[(f_1)_S(c(t)),(f_2)_S(c(t+\lambda))\right]^{(f_2)_T(c(t+\lambda))}-\left[(f_1)_S(c(t)),(f_2)_S(c(t))\right]^{(f_2)_T(c(t))}=
$$
$$
=\left[(f_1)_S(c(t)),(f_2)_S(c(t))+D((f_2)_S\circ c)(t)\lambda +\right.
$$
$$
\left.+\frac{1}{2}\lambda ^2\sum\limits_{i=1}^{n-1}((f_2)_S\circ c)''_i(t_i)e_i\right]^{(f_2)_T(c(t+\lambda))}-\left[(f_1)_S(c(t)),(f_2)_S(c(t))\right]^{(f_2)_T(c(t))}=
$$
$$
=\left(\left[(f_1)_S(c(t)),(f_2)_S(c(t))+D((f_2)_S\circ c)(t)\lambda \right]^{(f_2)_T(c(t))}-\right.
$$
$$
\left.-\left[(f_1)_S(c(t)),(f_2)_S(c(t))\right]^{(f_2)_T(c(t))}\right)+\left(\left[(f_1)_S(c(t)),(f_2)_S(c(t)) +\right. \right.
$$
$$
\left.+D((f_2)_S\circ c)(t)\lambda+\frac{1}{2}\lambda ^2\sum\limits_{i=1}^{n-1}((f_2)_S\circ c)''_i(t_i)e_i \right]^{(f_2)_T(c(t+\lambda))}-
$$
$$
\left.-\left[(f_1)_S(c(t)),(f_2)_S(c(t))+D((f_2)_S\circ c)(t)\lambda \right]^{(f_2)_T(c(t))}\right).
$$
Dividing by $\lambda $ and applying the limit procedure when $\lambda $ tends to zero we get from the first bracket the value:
$$
\left([(f_1)_S(c(t)),\cdot]^{(f_2)_T(c(t))}\right)'_{D((f_2)_S\circ{c})(t)}(((f_2)_S\circ c)(t))).
$$
We determine the value of the second bracket. By Definition 10 the second term in this bracket is
$$
\left[(f_1)_S(c(t)),(f_2)_S(c(t))+D((f_2)_S\circ c)(t)\lambda \right]^{(f_2)_T(c(t))}=
$$
$$
=\left[(f_1)_S(c(t)),(f_2)_S(c(t))+D((f_2)_S\circ c)(t)\lambda \right]^{(f_2)_T(c(t+\lambda))}-
$$
$$
-\frac{\partial{\left[(f_1)_S(c(t)),(f_2)_S(c(t))+D((f_2)_S\circ c)(t)\lambda \right]^{(f_2)_T(c(t))}}}{\partial \tau}\lambda'-o(\lambda'),
$$
where
$$
(f_2)_T(c(t+\lambda))=(f_2)_T(c(t))+\lambda' \mbox{ and }
\lim \limits_{\lambda'\mapsto 0}\frac{o(\lambda')}{\lambda'}=0.
$$
Since
$$
(f_2)_Tc(t+\lambda)=(f_2)_Tc(t)+\lambda\left((f_2)_T\circ c\right)'(t) +o_1(\lambda),
$$
we have that
$$
\lambda' =\lambda\left((f_2)_T\circ c\right)'(t) +o_1(\lambda).
$$
By the Lipschitz condition (which also holds in the second variable of the product) there is a real constant $K$ with which we have that the absolute value of the substraction
$$
\left(\left[(f_1)_S(c(t)),(f_2)_S(c(t)) +\right. \right.
$$
$$
\left.+D((f_2)_S\circ c)(t)\lambda+\frac{1}{2}\lambda ^2\sum\limits_{i=1}^{n-1}((f_2)_S\circ c)''_i(t_i)e_i \right]^{(f_2)_T(c(t+\lambda))}-
$$
$$
-\left[(f_1)_S(c(t)),(f_2)_S(c(t))+D((f_2)_S\circ c)(t)\lambda \right]^{(f_2)_T(c(t+\lambda))}
$$
is less than or equal to
$$
K\left[(f_1)_S(c(t)),\frac{1}{2}\lambda ^2\sum\limits_{i=1}^{n-1}((f_2)_S\circ c)''_i(t_i)e_i\right]^{(f_2)_T(c(t+\lambda))}.
$$
Dividing by $\lambda $ and applying the limit procedure as $\lambda \rightarrow 0$, this quantity tends to zero. Dividing also by $\lambda $, for the remaining parts we have
$$
\frac{1}{\lambda}\frac{\partial{\left[(f_1)_S(c(t)),(f_2)_S(c(t))+D((f_2)_S\circ c)(t)\lambda \right]^{(f_2)_T(c(t))}}}{\partial \tau}\lambda'+o(\lambda')=
$$
$$
=\frac{\partial{\left[(f_1)_S(c(t)),(f_2)_S(c(t))+\lambda D((f_2)_S\circ c)(t)\right]^{(f_2)_T(c(t))}}}{\partial \tau}\times 
$$
$$
\times \left(\left((f_2)_T\circ c\right)'(t) +\frac{o_1(\lambda)}{\lambda}\right)+\left(\frac{o\left(\lambda\left((f_2)_T\circ c\right)'(t) +o_1(\lambda)\right)}{\lambda\left((f_2)_T\circ c\right)'(t) +o_1(\lambda)}\right)\times
$$
$$
\times \left(\frac{\lambda\left((f_2)_T\circ c\right)'(t) +o_1(\lambda)}{\lambda}\right),
$$
and if $\lambda$ tends  to zero then it is equal to
$$
\frac{\partial{\left[(f_1)_S(c(t)),(f_2)_S(c(t)) \right]^{\tau}}}{\partial \tau}((f_2)_T(c(t)))\cdot((f_2)_T\circ c)'(t).
$$
Thus, we proved our statement on the space-like component. On the other hand $(f_1)_T$, $(f_2)_T$,
are real-real functions, respectively. This implies that
$$
\lim\limits_{\lambda \rightarrow 0}\frac{1}{\lambda}([(f_1)_T(c(t+\lambda )),(f_2)_T(c(t+\lambda ))]-[(f_1)_T(c(t)),(f_2)_T(c(t))])=
$$
$$
=-((f_1)_T\circ c)'(t)(f_2)_T(c(t)) -(f_1)_T(c(t))((f_2)_T\circ c)'(t)
$$
showing the assertion of the theorem.
\qed

Let $F$ be a hypersurface defined by the function $f:S\longrightarrow V=S+T$. Here
$$
f(s)=s+\mathfrak{f}(s)e_n
$$ denotes points of $F$. The $C^2$ curve $c:\mathbb{R}\longrightarrow S$ define  a curve on $F$.
We collect the most important formulas of time-space in a list.
\begin{itemize}
\item  The \emph{first fundamental form} at the point $(f(c(t))$ of the hypersurface $F$  is the product
$$
\mathrm I_{f(c(t)}:=[D(f\circ c)(t),D(f\circ c)(t)]^{+,T}=
$$
$$
=[\dot{c}(t),\dot{c}(t)]^{\mathfrak{f}(c(t))}-[(\mathfrak{f}\circ c)'(t)]^2.
$$
\item The \emph{second fundamental form} at the point $f(c(t))$ is:
$$
\mathrm I \mathrm I:=[D^2(f\circ c)(t),(n^0\circ c)(t)]^{+,T}_{(f\circ c)(t)}=
$$
$$
=-{\left([D(f\circ c)(t),\cdot]^{+,T}\right)}'_{D(n^0\circ c)(t)}((n^0\circ c)(t))-
$$
$$
-\frac{\partial\left([\dot{c}(t),c(t)]^{\tau}\right)}{\partial \tau}(\mathfrak{n}^0(c(t)))\cdot(\mathfrak{n}^0\circ c)'(t),
$$
where $n^0$ is the unit normal vector defined from a normal vector $n(s)=s+\mathfrak{n}(s)e_n$ by
$$
n^0(c(t)):=\left\{\begin{array}{cc}n(c(t)) & \mbox{ if } n \mbox{ light-like vector}\\
\frac{n(c(t))} {\sqrt{|[n(c(t)),n(c(t))]^{+,T}|}} & \mbox{ otherwise. }\end{array}\right.
$$
\item If we consider a two-plane in the tangent hyperplane at $f(c(t))$ then it has a two dimensional pre-image in $(S,\|\cdot\|^{\mathfrak{f}(c(t))})$ by the regular linear mapping $Df$. In this plane we can consider an Auerbach basis $\{e_1,e_2\}$. The \emph{sectional principal curvatures} of a 2-section of the tangent hyperplane in the direction of the 2-plane spanned by $\{u=Df(e_1)$ and $v=Df(e_2)\}$ are the extremal values of the function
$$
\rho (D(f\circ c)):=\frac{\mathrm I \mathrm I_{f\circ c(t)}}{\mathrm I_{f\circ c(t)}},
$$
of the variable $D(f\circ c)$. We denote them by $\rho(u,v)_{\mathrm{max}}$ and $\rho(u,v)_{\mathrm{min}}$, respectively. The \emph{sectional (Gauss) curvature} $\kappa(u,v)$  (at the examined point $c(t)$) is the product
$$
\kappa(u,v):=[n^0(c(t)),n^0(c(t))]^{+,T} \rho (u,v)_{\mathrm{max}}\rho(u,v)_{\mathrm{min}}.
$$
\item The \emph{Ricci curvature} $\mathrm{Ric}(v)$ in the direction of the tangent vector $v$ at the point $f(c(t))$ is
$$
\mathrm{Ric}(v)_{f(c(t))} := (n - 2)\cdot E(\kappa_{f(c(t))}(u,v))
$$
where $\kappa_{f(c(t))}(u,v)$ is the random variable of the sectional curvatures of the two planes spanned by $v$ and a random  $u$ of the tangent hyperplane holding the equality $[u,v]^{+,T}=[u,v]^{\mathfrak{f}(c(t))}=0$. We also say that the scalar curvature of the hypersurface $f$ at its point $f(c(t))$ is
$$
\Gamma_{f(c(t))}:= {n-1 \choose 2} \cdot E(\kappa_{f(c(t))}(u,v)).
$$
\end{itemize}

\subsection{Surfaces defined by implicite functions}

In this section we investigate such sets of a deterministic time-space model which are not hypersurfaces. The importance of this examination in that the most nice subsets as the imaginary unit sphere or the de Sitter sphere belong to this class. On the other hand, it can be observed that assuming certain smoothness condition these sets handled locally as hypersurfaces and can also determine their differential geometric properties.

\subsubsection{Imaginary unit sphere}

The points of $H^{+,T}$ can be defined by the union
$$
\cup\left\{\left\{s+\tau \mbox{ where } \sqrt{[s,s]^\tau+1}=\tau \right\} \mbox{ , } \tau\geq 1\right\}.
$$
Our assumption on $K(\tau)$  cannot guaranty  that for every $s\in S$ there is at least one $\tau $ is holding the equality $\sqrt{[s,s]^\tau+1}=\tau$. On the other hand if we assume that $\rho_H(K(\tau),B_E)\leq 1$ the ball $2K(\tau)$ contains the Euclidean ball $B_E$ for every $\tau$. Hence $[s,s]^\tau \leq 4\|s\|_E^2$ so for all $\tau$ with $\tau ^2 > 4\|s\|_E^2+1$, the inequality $[s,s]^\tau +1<\tau ^2$ holds. Since for a non-zero vector $s$ we have $[s,s]^1 +1>1$, the statement follows by continuity variable. In the non-trivial case the sets defined by distinct moments have distinct shape. From this immediately follows that $H^{+,T}$ is not a hypersurface of the time-space hence its differential geometry can be considered only on the base of its implicit definition. Consider the function $\mathfrak{H}:V\rightarrow \mathbb{R}$ defined by
$$
\mathfrak{H}(s+\tau e_n):=\sqrt{[s,s]^\tau+1}-\tau.
$$
If $v_0=s_0+\tau_0 e_n$ is a point on $H^{+,T}$ then we have $\mathfrak{H}(v_0)=0$. By our definition $\mathfrak{H}$ is continuously differentiable at the point $v_0$. Assume that
$$
\frac{\partial \mathfrak{H}}{\partial \tau}(v_0)= \frac{\frac{\partial([s,s]^\tau)}{\partial \tau}}{2\sqrt{[s,s]^\tau+1}}(v_0)-1\neq 0,
$$
or equivalently
$$
\frac{\partial([s_0,s_0]^\tau)}{\partial \tau}(\tau_0)\neq 2\sqrt{[s_0,s_0]^{\tau_0}+1}.
$$
Then by implicit function theorem there is a neighborhood $U$ of $v_0$ and a function $\mathfrak{h}: S\rightarrow \mathbb{R}$ such that $\tau=\mathfrak{h}(s)$ for the points $v=s+\tau e_n$ of $H^{+,T}$. Thus we have in $U$ (as in Lemma 3 in \cite{gho2}) that
$$
\mathfrak{h}(s)=\sqrt{[s,s]^{\mathfrak{h}(s)}+1}.
$$

If the vector $s$ comes from a point of a curve $c(t)\subset S$ by the definition $c(t+\lambda)=s+\lambda e$, we get the equalities:
$$
(\mathfrak{h}\circ c)(t)=\sqrt{[(c(t),c(t)]^{\mathfrak{h}(c(t))}+1}
$$
and also
$$
(\mathfrak{h}\circ c)'(t)=\frac{[\dot{c}(t),c(t)]^{\mathfrak{h}(c(t))}}{\sqrt{1+[c(t),c(t)]^{\mathfrak{h}(c(t))}}}+
\frac{\frac{\partial{\left[c(t),c(t) \right]^{\tau}}}{\partial \tau}(\mathfrak{h}(c(t)))\cdot(\mathfrak{h}\circ c)'(t)}{2\sqrt{1+[c(t),c(t)]^{\mathfrak{h}(c(t))}}}
$$
or equivalently,
$$
(\mathfrak{h}\circ c)'(t)=\left(1-\frac{\frac{\partial{\left[c(t),c(t) \right]^{\tau}}}{\partial \tau}(\mathfrak{h}(c(t)))}{2\sqrt{1+[c(t),c(t)]^{\mathfrak{h}(c(t))}}}\right)^{-1} \frac{[\dot{c}(t),c(t)]^{\mathfrak{h}(c(t))}}{\sqrt{1+[c(t),c(t)]^{\mathfrak{h}(c(t))}}}=
$$
$$
=\frac{2}{2\mathfrak{h}(c(t))-\frac{\partial{\left[c(t),c(t) \right]^{\tau}}}{\partial \tau}(\mathfrak{h}(c(t)))}[\dot{c}(t),c(t)]^{\mathfrak{h}(c(t))}
$$
as on page 25. in \cite{gho2}. We note that the additional value
$$
\frac{\partial{\left[c(t),c(t) \right]^{\tau}}}{\partial \tau}(\mathfrak{h}(c(t)))
$$
of the formula depend on the position $c(t+0)=s$ and do not depend on the direction $e$.
Thus the first fundamental form is:
$$
\mathrm I=[\dot{c}(t)+(\mathfrak{h}\circ c)'(t)e_n, \dot{c}(t)+(\mathfrak{h}\circ c)'(t)e_n]^{+,T}=
$$
$$
=[\dot{c}(t),\dot{c}(t)]^{(\mathfrak{h}\circ c)'(t)}-[(\mathfrak{h}\circ c)'(t)]^2=
$$
$$
=[\dot{c},\dot{c}]^{\frac{2[\dot{c}(t),c(t)]^{\mathfrak{h}(c(t))}}{2\mathfrak{h}(c(t))-\frac{\partial{\left[c(t),c(t) \right]^{\tau}}}{\partial \tau}(\mathfrak{h}(c(t)))}}-\left(\frac{2[\dot{c}(t),c(t)]^{\mathfrak{h}(c(t))}}{2\mathfrak{h}(c(t))-\frac{\partial{\left[c(t),c(t) \right]^{\tau}}}{\partial \tau}(\mathfrak{h}(c(t)))}\right)^{2}.
$$
To calculate the second fundamental form we have to determine the unit normal vector field. A tangent vector is
$$
\dot{c}(t)+(\mathfrak{h}\circ c)'(t)e_n=\dot{c}(t)+
$$
$$
+\left(1-\frac{\frac{\partial{\left[c(t),c(t) \right]^{\tau}}}{\partial \tau}(\mathfrak{h}(c(t)))}{2\sqrt{1+[c(t),c(t)]^{\mathfrak{h}(c(t))}}}\right)^{-1} \frac{[\dot{c}(t),c(t)]^{\mathfrak{h}(c(t))}}{\sqrt{1+[c(t),c(t)]^{\mathfrak{h}(c(t))}}}e_n.
$$
We may see that
$$
\left[\dot{c}(t)+\frac{2[\dot{c}(t),c(t)]^{\mathfrak{h}(c(t))}}{2\mathfrak{h}(c(t))-\frac{\partial{\left[c(t),c(t) \right]^{\tau}}}{\partial \tau}(\mathfrak{h}(c(t)))}e_n,\right.
$$
$$
\left. ,\frac{2{\mathfrak{h}(c(t))}}{2\mathfrak{h}(c(t))-\frac{\partial{\left[c(t),c(t) \right]^{\tau}}}{\partial \tau}(\mathfrak{h}(c(t)))}c(t)+\mathfrak{h}(c(t))e_n\right]^{+,T}=0
$$
showing the equality
$$
n\circ c=\frac{2{\mathfrak{h}(c(t))}}{2\mathfrak{h}(c(t))-\frac{\partial{\left[c(t),c(t) \right]^{\tau}}}{\partial \tau}(\mathfrak{h}(c(t)))}c(t)+(\mathfrak{h}\circ c)(t)e_n.
$$
The second fundamental form of $H^{+,T}$ is
$$
\mathrm I \mathrm I:=\left[\ddot{c}(t)+(\mathfrak{h}\circ c)''(t)e_n,\right.
$$
$$
\left. ,\frac{2{\mathfrak{h}(c(t))}}{2\mathfrak{h}(c(t))-\frac{\partial{\left[c(t),c(t) \right]^{\tau}}}{\partial \tau}(\mathfrak{h}(c(t)))}c(t)+(\mathfrak{h}\circ c)(t)e_n\right]^{+,T}_{(\mathfrak{h}\circ c)(t)}=
$$
$$
=\frac{2{\mathfrak{h}(c(t))}}{2\mathfrak{h}(c(t))-\frac{\partial{\left[c(t),c(t) \right]^{\tau}}}{\partial \tau}(\mathfrak{h}(c(t)))}[\ddot{c}(t),c(t)]^{(\mathfrak{h}\circ c)(t)}-(\mathfrak{h}\circ c)''(t)\mathfrak{h}(c(t)).
$$
We use here Theorem 1. Thus we get first that
$$
(\mathfrak{h}\circ c)''(t)=\left(\frac{2\left[\dot{c}(t),c(t)\right]^{\mathfrak{h}(c(t))}}{2\mathfrak{h}(c(t))-\frac{\partial[c(t),c(t)]^\tau}{\partial \tau}(\mathfrak{h}(c(t)))}\right)'=A+B
$$
where
$$
A=\frac{2}{2\mathfrak{h}(c(t))-\frac{\partial[c(t),c(t)]^\tau}{\partial \tau}(\mathfrak{h}(c(t)))}\left([\ddot{c}(t),c(t)]^{\mathfrak{h}(c(t))}+\right.
$$
$$
+\left.\left([\dot{c}(t),\cdot]^{\mathfrak{h}(c(t))}\right)'_{\dot{c}(t)}(c(t))+\frac{2[\dot{c}(t),c(t)]^{\mathfrak{h}(c(t))}\frac{\partial[\dot{c}(t),c(t)]^\tau}{\partial \tau}(\mathfrak{h}(c(t)))}{2\mathfrak{h}(c(t))-\frac{\partial[c(t),c(t)]^\tau}{\partial \tau}(\mathfrak{h}(c(t)))}\right)
$$
and
$$
B=\frac{-2[\dot{c}(t),c(t)]^{\mathfrak{h}(c(t))}}{\left(2\mathfrak{h}(c(t))-\frac{\partial[c(t),c(t)]^\tau}{\partial \tau}(\mathfrak{h}(c(t)))\right)^2}\times 
$$
$$
\times 2\left((\mathfrak{h}\circ c)'(t)\left(1-\frac{1}{2}\frac{\partial^2 [{c}(t),c(t)]^\tau}{\left(\partial \tau\right)^2}(\mathfrak{h}(c(t)))\right)-\frac{\partial [\dot{c}(t),c(t)]^\tau}{\partial \tau}(\mathfrak{h}(c(t)))\right)=
$$
$$
=\frac{-2}{2\mathfrak{h}(c(t))-\frac{\partial[c(t),c(t)]^\tau}{\partial \tau}(\mathfrak{h}(c(t)))}\times 
$$
$$
\times \left(\left((\mathfrak{h}\circ c)'(t)\right)^2\left(1-\frac{1}{2}\frac{\partial^2 [{c}(t),c(t)]^\tau}{\left(\partial \tau\right)^2}(\mathfrak{h}(c(t)))\right)-\right.
$$
$$
\left.-(\mathfrak{h}\circ c)'(t)\frac{\partial [\dot{c}(t),c(t)]^\tau}{\partial \tau}(\mathfrak{h}(c(t)))\right).
$$
Since in time-space model the result of Lemma 3 of paper \cite{gho2} of the generalized space-time model can be interpreted as
$$
\left([\dot{c}(t),\cdot]^{\mathfrak{h}(c(t))}\right)'_{\dot{c}(t)}(c(t))=[\dot{c}(t),\dot{c}(t)]^{(\mathfrak{h}\circ c)'(t))}
$$
we get that the second fundamental form is:
$$
\mathrm I \mathrm I=\frac{2{\mathfrak{h}(c(t))}}{2\mathfrak{h}(c(t))-\frac{\partial{\left[c(t),c(t) \right]^{\tau}}}{\partial \tau}(\mathfrak{h}(c(t)))}[\ddot{c}(t),c(t)]^{\mathfrak{h}(c(t))}-(\mathfrak{h}\circ c)''(t)\mathfrak{h}(c(t))=
$$
$$
=\frac{2{\mathfrak{h}(c(t))}}{2\mathfrak{h}(c(t))-\frac{\partial{\left[c(t),c(t) \right]^{\tau}}}{\partial \tau}(\mathfrak{h}(c(t)))}\left[-[\dot{c}(t),\dot{c}(t)]^{(\mathfrak{h}\circ c)'(t)}+\right.
$$
$$
+\left((\mathfrak{h}\circ c)'(t)\right)^2\left(1-\frac{1}{2}\frac{\partial^2 [{c}(t),c(t)]^\tau}{\left(\partial \tau\right)^2}(\mathfrak{h}(c(t)))\right)-
$$
$$
\left.-2(\mathfrak{h}\circ c)'(t)\frac{\partial [\dot{c}(t),c(t)]^\tau}{\partial \tau}(\mathfrak{h}(c(t)))\right],
$$
where
$$
(\mathfrak{h}\circ c)'(t))=\frac{2[\dot{c}(t),c(t)]^{\mathfrak{h}(c(t))}}{2\mathfrak{h}(c(t))-\frac{\partial{\left[c(t),c(t) \right]^{\tau}}}{\partial \tau}(\mathfrak{h}(c(t)))}.
$$

Observe, that if the norm is a constant function of the time, these formulas simplify to the formulas of the generalized space-time model. We now give examples to illustrate that this basic tools of the corresponding differential geometry can be calculated.

\begin{examples}
\begin{enumerate}

\item For a $3$-dimensional example we take the function $K(\tau):\tau\mapsto G_{\tau}$, where $G_{\tau}$ is the ellipse of area $\pi$ with half-axes  $\tau e_1$ and $\frac{1}{\tau}e_2$. Here $\{e_1,e_2\}$ is an orthonormed basis of the embedding Euclidean plane. The connection between the norms of the vector $s=xe_1+ye_2$ and its Euclidean coordinates is
$$
[s,s]^{\tau}=\tau ^2x^2+\frac{y^2}{\tau^2}.
$$
The implicite equation for the corresponding imaginary unit sphere is
$$
\tau=\sqrt{1+\tau ^2x^2+\frac{y^2}{\tau^2}},
$$
if we assume that
$$
2\tau x^2-\frac{2y^2}{\tau ^3}\neq 2\tau,
$$
or equivalently
$$
x^2-1\neq \frac{y^2}{\tau^4}.
$$
For a vector $s=(x,y)^T$ we exclude the time $\tau $ with equality
$$
\tau ^4=\frac{y^2}{x^2-1}
$$
where $x^2\neq 1$. (Thus if $x^2=1$ there is no $\tau $, which need to exclude from the investigation.)
Solving the implicite equation we get that
$$
\tau^2=\frac{1\pm \sqrt{1+4(1-x^2)y^2}}{2(1-x^2)} \mbox{ if } x^2\neq 1,
$$
and in the case when $x^2=1$ $\tau $ has to be $\infty $ for every $y$.
This formula shows that we can get real values for $\tau $ if and only if
$$
x^2\leq 1+\frac{1}{4y^2}.
$$
Thus the domain of the imaginary unit sphere is the union of three domains bounded by the curves $x=\pm 1$ and $x=\pm \sqrt{1+\frac{1}{4y^2}}$ drawing on the figure Fig.2.

\begin{figure}[htbp]
\centerline{\includegraphics[scale=1]{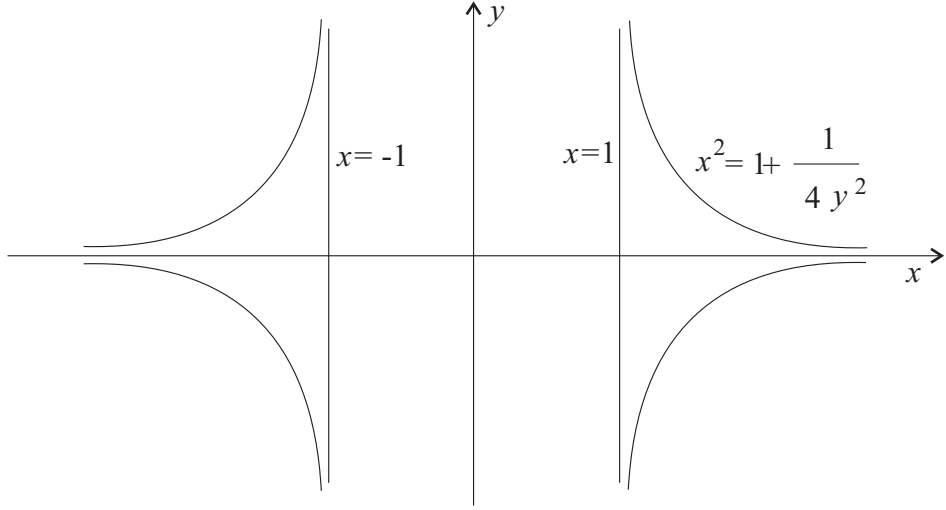}}
\caption{The domain of the imaginary unit sphere in the example.}
\end{figure}

Since $\tau ^2>0$ we also have that if $|x|<1$ then we have to consider the equality with positive sign
$$
\tau^2=\frac{1+\sqrt{1+4(1-x^2)y^2}}{2(1-x^2)},
$$
and for the other two connected components we have to choose the equality with negative sign:
$$
\tau^2=\frac{1- \sqrt{1+4(1-x^2)y^2}}{2(1-x^2)}.
$$
The first fundamental form is
$$
\mathrm{ I }=[\dot{c},\dot{c}]^{\frac{2[\dot{c}(t),c(t)]^{\mathfrak{h}(c(t))}}{2\mathfrak{h}(c(t))-\frac{\partial{\left[c(t),c(t) \right]^{\tau}}}{\partial \tau}(\mathfrak{h}(c(t)))}}-\left(\frac{2[\dot{c}(t),c(t)]^{\mathfrak{h}(c(t))}}{2\mathfrak{h}(c(t))-\frac{\partial{\left[c(t),c(t) \right]^{\tau}}}{\partial \tau}(\mathfrak{h}(c(t)))}\right)^{2}.
$$
Since
$$
[\dot{c}(t),c(t)]^{\mathfrak{h}(c(t))}=\mathfrak{h}(c(t))^2\dot{x(t)}x(t)+\frac{\dot{y(t)}y(t)}{\mathfrak{h}(c(t))^2},
$$
$$
\frac{\partial{\left[c(t),c(t) \right]^{\tau}}}{\partial \tau}(\mathfrak{h}(c(t)))=2\mathfrak{h}(c(t)) x(t)^2-\frac{2y(t)^2}{\mathfrak{h}(c(t))^3},
$$
we have that
$$
\mathrm{ I }=((\mathfrak{h}\circ c)'(t))^2\left((\dot{x}(t))^2-1\right)+\frac{(\dot{y}(t))^2}{((\mathfrak{h}\circ c)'(t))^2},
$$
where
$$
(\mathfrak{h}\circ c)'(t)=\mathfrak{h}(c(t)\frac{\left(\mathfrak{h}(c(t)\right)^4\dot{x}(t)x(t)+\dot{y}(t)y(t)}{\left(\mathfrak{h}(c(t)\right)^4\left(1-(x(t))^2\right)+(y(t))^2}
$$
with
$$
\left(\mathfrak{h}(c(t))\right)^2=\frac{1\pm \sqrt{1+4(1-(x(t))^2)(y(t))^2}}{2(1-(x(t))^2)}.
$$
We also get that
$$
\mathrm{ I }\mathrm{ I }=-\frac{2{\mathfrak{h}(c(t))}}{2\mathfrak{h}(c(t))-\frac{\partial{\left[c(t),c(t) \right]^{\tau}}}{\partial \tau}(\mathfrak{h}(c(t)))}\left[((\mathfrak{h}\circ c)'(t))^2(\dot{x}(t))^2+\right.
$$
$$
+\frac{(\dot{y}(t))^2}{((\mathfrak{h}\circ c)'(t))^2}-((\mathfrak{h}\circ c)'(t))^2\left(1-\frac{1}{2}\frac{\partial^2[c(t),c(t)]^\tau}{\partial \tau}(\mathfrak{h}(c(t)))\right)+
$$
$$
+\left.2(\mathfrak{h}\circ c)'(t)\frac{\partial [\dot{c}(t),c(t)]^\tau}{\partial \tau}(\mathfrak{h}(c(t)))\right]=
$$
$$
=-\frac{2{\mathfrak{h}(c(t))}}{2\mathfrak{h}(c(t))-\frac{\partial{\left[c(t),c(t) \right]^{\tau}}}{\partial \tau}(\mathfrak{h}(c(t)))}\left[((\mathfrak{h}\circ c)'(t))^2\left((\dot{x}(t))^2-1+\right.\right.
$$
$$
\left.+(x(t))^2+\frac{3(y(t))^2}{\left(\mathfrak{h}(c(t))\right)^4}\right)+
$$
$$
\left.+4(\mathfrak{h}\circ c)'(t)\left(\mathfrak{h}(c(t))\dot{x}(t)x(t)-\frac{\dot{y}(t)y(t)}{\left(\mathfrak{h}(c(t))\right)^3}\right)+\frac{\left(\dot{y}(t)\right)^2} {\left((\mathfrak{h}\circ c)'(t)\right)^2}\right].
$$
For a more concrete example assume that
$$
c(t)=(x(t),y(t))=(t\cos \alpha, \sqrt{2}+t\sin \alpha), \mbox{ and } t_0=0.
$$

Then we have that $\left(\mathfrak{h}(c(t_0))\right)^2=2$ because in the formula
$$
\frac{1\pm \sqrt{1+4(1-x(t)^2)y(t)^2}}{2(1-x(t)^2)}
$$
we have to calculate with positive sign. Since
$$
(\mathfrak{h}\circ c)'(t_0)=\sqrt{2}\frac{\sqrt{2}\sin \alpha}{4+2}=\frac{1}{3}\sin \alpha,
$$ we get that
$$
\mathrm I =\frac{1}{9}\sin ^2\alpha(\cos ^2\alpha -1)+\frac{\sin^2\alpha}{\frac{1}{9}\sin^2\alpha}=9-\frac{1}{9}\sin ^4\alpha> 0.
$$
Similarly the second fundamental form is
$$
\mathrm I \mathrm I=-\frac{2}{3}\left(\frac{1}{9}\sin ^2\alpha\left(\cos ^2\alpha -1+ \frac{3}{2}\right)+9+\frac{2\sqrt{2}}{3}\sin^2\alpha\right)=
$$
$$
=-\frac{2}{3}\left(\left(\frac{1}{6}+\frac{2\sqrt{2}}{3}\right)\sin ^2\alpha-\frac{1}{9}\sin^4\alpha+9\right)=
$$
$$
=-\frac{1+4\sqrt{2}}{9}\sin ^2\alpha +\frac{2}{27}\sin ^4\alpha-6.
$$

The extremal values of the non-positive function
$$
\frac{\mathrm I \mathrm I}{\mathrm I }=\frac{\frac{2}{27}\sin ^4\alpha -\frac{1+4\sqrt{2}}{9}\sin ^2\alpha-6}{9-\frac{1}{9}\sin ^4\alpha}
$$
attained at the directions $\alpha$ for which either $\cos \alpha =0$ or $\sin\alpha =0$ with the respective negative values $-\frac{157+12\sqrt{2}}{240}$ and $-\frac{2}{3}$. Since the normal vector at this point is
$$
n\circ c=\frac{2{\mathfrak{h}(c(t))}}{2\mathfrak{h}(c(t))-\frac{\partial{\left[c(t),c(t) \right]^{\tau}}}{\partial \tau}(\mathfrak{h}(c(t)))}c(t)+(\mathfrak{h}\circ c)(t)e_3=
$$
$$
=\frac{2}{3}(0,\sqrt{2})^T+\sqrt{2}e_3=\sqrt{2}\left(\left(0,\frac{2}{3}\right)^T+e_3\right),
$$
we have that the norm-square of it is
$
2\left(\frac{2}{9}-1\right)=-\frac{14}{9}<0
$
and hence the Gaussian curvature is negative at this point.

\item For a further example choose an ellipse $G_{\alpha}$ as in the previous example with a fixed parameter $\alpha$, where $1\leq\alpha\leq 2$. Let $K(\tau)$ be the rotated copy of this ellipse about the time axis with the angle $\tau$. Then
    $$
    [s,s]^{\tau}=[xe_1+ye_2,xe_1+ye_2]^{\tau}=
    $$
    $$
    =\alpha ^2(\cos \tau x +\sin \tau y)^2+\frac{(-\sin \tau x+\cos \tau y)^2}{\alpha ^2}=
    $$
    $$
    =\left(\alpha ^2 x^2+\frac{y^2}{\alpha ^2}\right)\cos^2\tau+\left(\alpha ^2 y^2+\frac{x^2}{\alpha ^2}\right)\sin^2\tau+2\cos \tau\sin\tau \left(\alpha ^2-\frac{1}{\alpha ^2}\right)=
    $$
    $$
    =\left(\alpha ^2 x^2+\frac{y^2}{\alpha ^2}\right)+\left(\alpha ^2-\frac{1}{\alpha ^2}\right)\left(y^2-x^2\right)\sin^2\tau +2\cos \tau\sin\tau \left(\alpha ^2-\frac{1}{\alpha ^2}\right)=
    $$
    $$
    =\left(\alpha ^2 x^2+\frac{y^2}{\alpha ^2}\right)+\left(\alpha ^2-\frac{1}{\alpha ^2}\right)\left(y^2-x^2\right)\frac{1}{2}-
    $$
    $$
    -\frac{1}{2}\left(\alpha ^2-\frac{1}{\alpha ^2}\right)\left(y^2-x^2\right)\cos2\tau+\sin2\tau\left(\alpha ^2-\frac{1}{\alpha ^2}\right)=
    $$
    $$
    =\frac{1}{2}(\alpha^2+\frac{1}{\alpha^2})(x^2+y^2)+\left(\alpha ^2-\frac{1}{\alpha ^2}\right)\left(\sin2\tau -\frac{1}{2}\left(y^2-x^2\right)\cos2\tau\right)
    $$
The implicite equation of the imaginary unit sphere is
$$
\tau=\sqrt{1+\frac{\alpha^4+1}{2\alpha^2}(x^2+y^2)+\frac{\alpha^4-1}{\alpha ^2}\left(\sin2\tau -\frac{1}{2}\left(y^2-x^2\right)\cos2\tau\right)}.
$$
Here there is no explicite form for $\tau$ however in a concrete point the fundamental forms and curvatures can be determined. We remark that the Hausdorff distances of the unit ball $K(\tau)$ from $B_E$ is less or equal to 1, thus the domain is the whole plane. Since the norm induced by an inner product in every time the corresponding time-space is a semi-Riemann manifold.

\item We can get premanifolds if the square of the examined norms can not be represented as the scalar square of an inner product. A three-dimensional example can be gotten from the function $K(\tau)$ which sends $\tau$ for $\tau> 1$ to the unit ball of the $l_{\tau}$ space with Euclidean area $\pi$. In this case
    $$
    [s,s]^{\tau}=\frac{v(l_\tau)}{\pi}\sqrt[\tau]{|x|^{\tau}+|y|^{\tau}},
    $$
where
$$
v(l_\tau)=\frac{\Gamma\left(1+\frac{1}{\tau}\right)^2}{\Gamma\left(1+\frac{2}{\tau}\right)}4
$$
is the volume of the unit ball of the standard $l_\tau$ norm of the plane. Here for $\tau $ we have the implicite equality
$$
\tau=\sqrt{1+\frac{v(l_\tau)}{\pi}\sqrt[\tau]{|x|^{\tau}+|y|^{\tau}}}.
$$
As in the previous example the domain is also the plane $S$.

\end{enumerate}
\end{examples}

\subsubsection{de Sitter sphere}

The points of the de Sitter sphere $G^{+,T}$ can be defined by the union
$$
\cup\left\{\left\{s+\tau e_n \mbox{ where } \sqrt{[s,s]^\tau-1}=\tau \right\} \mbox{ , } [s,s]^\tau\geq 1\right\}.
$$
$G^+$ is not a hypersurface. It can be handled also by an implicite function
$$
\tau=\sqrt{-1+[s,s]^{\tau}} \mbox{ for } [s,s]^\tau>1,
$$
using the assumption
$$
\frac{\partial \mathfrak{G}}{\partial \tau}(v_0)= \frac{\frac{\partial([s,s]^\tau)}{\partial \tau}}{2\sqrt{[s,s]^\tau-1}}(v_0)-1\neq 0,
$$
or equivalently
$$
\frac{\partial([s_0,s_0]^\tau)}{\partial \tau}(\tau_0)\neq 2\sqrt{[s_0,s_0]^{\tau_0}-1}.
$$
Using the equality
$$
\mathfrak{h}^2(s)+\mathfrak{g}^2(s)=[s,s]^{\mathfrak{h}(s)}+[s,s]^{\mathfrak{g}(s)},
$$
the derivative of $\mathfrak{g}$ in the direction of the unit vector $e\in S$ can be calculated from the equality
$$
2\mathfrak{h}(s)\mathfrak{h}'_e(s)+2\mathfrak{g}(s)\mathfrak{g}'_e(s)=\left([s,s]^{\mathfrak{h}(s)}+[s,s]^{\mathfrak{g}(s)}\right)'=
$$
$$
=\left(2[e,s]^{\mathfrak{h}(s)}+ \frac{\partial{\left[s,s \right]^{\mathfrak{h}(s)}}}{\partial \tau}(\tau)\cdot \mathfrak{h}'_e(s)\right)+\left(2[e,s]^{\mathfrak{g}(s)}+ \frac{\partial{\left[s,s \right]^{\mathfrak{g}(s)}}}{\partial \tau}(\tau)\cdot \mathfrak{g}'_e(s)\right) .
$$
Thus
$$
\mathfrak{g}'_e(s)=\frac{2[e,s]^{\mathfrak{g}(s)}}{2\mathfrak{g}(s)-\frac{\partial{\left[s,s \right]^{\mathfrak{g}(s)}}}{\partial \tau}(\mathfrak{g}(s))}.
$$
The first and second fundamental forms have analogous forms as in the case of the imaginary unit sphere $H^{+,T}$.

\section{On the mechanics of the time-like vectors}

In this section we investigate the objects of the time-like vectors of a deterministic time-space model. We consider the upper part of this set restricting our investigation to the positive elements of $T$, denoted by this set $T^+$. The theory of generalized space-time model can be used in a generalization of special relativity theory, if we change some previous formulas using also the constant $c$. (It is practically can be considered as the speed of the light in vacuum.) The formula of the product in such a deterministic (random) time-space is
$$
[x',x'']^{+,T}:=[s',s'']^{\tau ''}+c^2\left[\tau ',\tau ''\right].
$$
Parallel we use the assumption that the dimension $n$ is equal to $4$.

\subsection{The word-line of a particle}

A particle is a random function $x: I_x \rightarrow S$ holding two conditions:
\begin{itemize}
\item the set $I_x\subset T^+$ is an interval
\item
$
[x(\tau),x(\tau)]^{\tau}<0 \mbox{ if } \tau\in I_x.
$
\end{itemize}
The particle lives on the interval $I_x$, is born at the moment $\inf I_x$ and dies at the moment $\sup I_x$. Since all time-sections of a time-space model is a normed space of dimension $n$ the Borel sets of the time-sections are independent from the time. This means that we can consider the physical specifies of a particle as a trajectory of a stochastic process. A particle ``realistic" if it holds the ``known laws of physic" and ``idealistic" otherwise. This is only a terminology for own use, the mathematical contain of the expression ``known laws of physics" is indeterminable. Since the norm (and thus the metric) in a time-space model changes by the time, the formulas of the density function of a fixed distribution also changes by the time. For example, if we say that both of the functions $f(x(\tau_1))$ and $f(x(\tau_2))$ have normal distribution on its domain $\tau_1K(\tau_1)$ and $\tau_2K(\tau_2)$ we have to use distinct formulas on their density functions, respectively. The uniform distribution is the only distribution which density function is independent from the time. First we introduce an inner metric $\delta_{K(\tau)}$ on the space at the moment $\tau$. We have two possibilities, either we can consider this space with its original metric
$$
\delta_{K (\tau)}(u,v):=\|u-v\|^\tau,
$$
(arise from the norm) -- at this time the space bounded and all distances are less then $2\tau$ -- or as another possibility we can define a distance which derives from the ball $\tau K(\tau)$ indirectly. For example let $u,v\in \tau K(\tau)$ be two points and denote by $(uv)_\infty$ and $(uv)_{-\infty}$ the intersection points of the line $(uv)$ and the boundary of the ball $\tau K(\tau)$, respectively. (Here the point $v$ separates the points $u$ and $(uv)_\infty$.) Let $\left(u,v,(uv)_\infty,(uv)_{-\infty}\right)$ denote the cross ratio of the four points and let
$$
\delta_{K (\tau)}(u,v):=\ln \left(u,v,(uv)_\infty,(uv)_{-\infty}\right)
$$
be the inner metric of the space $\tau K(\tau)$. We note that if the norm is Euclidean it is the usual distance of a modeled hyperbolic space (which is unbounded with respect to this metric).
These thread motivates the following definition:
\begin{defi}
Let $X(\tau):T\rightarrow \tau K(\tau)$ be a continuously differentiable (by the time) trajectory of the random function $\left(x(\tau)\mbox{ , }\tau\in I_x\right)$. We say that the particle $x(\tau)$ is \emph{realistic in its position} if for every $\tau\in I_x$ the random variable
$\delta_{K (\tau)}\left(X(\tau),x(\tau)\right)$ has normal distribution on $\tau K(\tau)$. In other words the stochastic process
$\left(\delta_{K (\tau)}\left(X(\tau),x(\tau)\right)\mbox{ , }\tau\in I_x\right)$ has stationary Gaussian process with respect to a given continuously differentiable function $X(\tau)$. We call the function $X(\tau)$ the \emph{world-line} of the particle $x(\tau)$.
\end{defi}
We note that the two metrics defined above are essentially agree for small distances, thus the concept of "realistic in its position" independent from the choice of $\delta_{K (\tau)}$. As a refinement of this concept we define another one, which can be considered as a generalization of the principle on the maximality of the speed of the light.
\begin{defi}
We say that a particle \emph{realistic in its speed} if it is realistic in its position and the derivatives of its world-line  $X(\tau)$ are time-like vectors.
\end{defi}
Since the shape of the sets of the time-like points in a time-space is not a cone, it is possible that $u$ is a time-like vector but $\alpha u$ is not with certain $\alpha $. On the other hand in a random time-space model the speed of those particles which realistic in its speed with a great probability are less than to the speed of the light. Note that our theory does not exclude the possibility of the existence of a particle which speed is greater to the speed of the light at a moment neither in the case of generalized space-time model or in the case of a particle which is realistic in its speed.

For such two particles $x',x''$ which are realistic in their position we can define a momentary distance by the equality:
$$
\delta(x'(\tau),x''(\tau))=\|X'(\tau)-X''(\tau)\|^{\tau}=\sqrt{[X'(\tau)-X''(\tau),X'(\tau)-X''(\tau)]^{+,T}}.
$$
We can say that two particles $x'$ and $x''$ are agree if the expected value of their distances is equal to zero. Let $I=I_{x'}\cap I_{x''}$ be the common part of their domains. The required equality is:
$$
E(\delta_{K(\tau)}(x'(\tau),x''(\tau)))=\int\limits_{I}\delta_{K(\tau)}(x'(\tau),x''(\tau))\mathrm{ d }\tau=
$$
$$
=\int\limits_{I}\|X'(\tau)-X''(\tau)\|^{\tau}\mathrm{ d }\tau=0.
$$

\subsection{Frames in time-space}

The first question is: How we define the so-called ``inertial frame" in our model? If we insist on ``a Descartes-system of the space which moving with a constant velocity" then we have to interpret two things; the concepts of Descartes system and the concept of velocity, respectively. In a deterministic time-space we have a function $K(\tau)$, and we have more possibilities to define orthogonality in a concrete moment $\tau$. We shall fixe a concept of orthogonality and we will consider it in every normed space. In the case when the norm induced by the Euclidean inner product this method should give the same result as the usual concept of orthogonality. The most natural choice is the concept of Birkhoff orthogonality (see in \cite{gho1}). Using it, in every normed space we can consider an Auerbach basis (see in \cite{gho1}) which can play the role of a basic coordinate frame. We can determine the coordinates of the point with respect to this basis.
We say that a frame is \emph{at rest with respect to the absolute time} if its origin (as a particle) is at rest with respect to the absolute time $\tau$ and the unit vectors of its axes are at rest with respect to a fixed Euclidean orthogonal basis of $S$. In this case the world line of the origin in the model is a vertical line (parallel to $T$); it is the collection of those points of the model which absolute space-coordinates do not changes by the change of the absolute time. Unfortunately, practically  we do not know an absolute coordinate system, and we can not check the immobility of the axes of such a frame. This motivates our definition on inertial frame and inertial frame ``at rest", respectively. We denote by $(S,\|\cdot\|^\tau)$ the normed space with unit ball $K(\tau)$. In $S$ we fix an Euclidean orthonormal basis and give the coordinates of a point (vector) of $S$ with respect to this basis. We get curves in $S$ parameterized by the time $\tau$. First we define the concept of a frame.

\begin{defi}
The system $\{f_1(\tau),f_2(\tau),f_3(\tau), o(\tau)\}\in (S,\|\cdot\|^{+\tau})\times \tau K(\tau)$ is a \emph{frame}, if
\begin{itemize}
\item $o(\tau)$ is a particle realistic in its speed,
with such a world-line
$$
O(\tau):T\rightarrow \tau K(\tau)
$$
which does not intersect the absolute time axis $T$,
\item the functions
$$
f_i(\tau):T\rightarrow \cup\left\{(S,\|\cdot\|^\tau) \mbox{ , } \tau\in T\right\}
$$
are continuously differentiable, for all fixed $\tau$,
\item
the system $\{f_1(\tau), f_2(\tau), f_3(\tau)\}$ is an Auerbach basis with origin $O(\tau)$ in the normed space $(S,\|\cdot\|^\tau)$.
\end{itemize}
\end{defi}

\begin{remark}
The condition that the frame building up from elements of an Auerbach basis is very strong. In the most cases the Auerbach basis is unique. In an inner product space a set of pairwise conjugate diameters of element $n$ of the unit ellipsoid gives an Auerbach basis. It is easy to see that every two Auerbach basis are isometric to each other, there is a linear isometry of the space sending the first into the second. Thus the set of the Auerbach bases can be gotten using the elements of the symmetry group of the space from a fixed one.  The following lemma is obviously and we leave its proof to the reader.

\begin{lemma}
For every $\varepsilon>0$ and a pair $\{K',\mathcal{A}'\}$ where $K'\in \mathcal{K}_0$ is a unit ball of $C^2$-class and $\mathcal{A}'$ is an Auerbach basis of the normed space $\left(S,\|\cdot\|_{K'}\right)$ there is a $\delta >0$ such that if for $K''$ holds $\delta_H(K',K'')<\delta $ then it can be found an Auerbach basis $\mathcal{A}''\in \left(S,\|\cdot\|_{K''}\right)$ for which $\delta_H(\mathcal{A}',\mathcal{A}'')<\varepsilon$ holds.
\end{lemma}
\end{remark}

Note, that for a good model we have to guarantee that Einstein's convention on the equivalence of the inertial frames can be remained for us. However at this time we have no possibility to give the concepts of "frame at rest" and the concept of "frame which moves constant velocity with respect to another one". The reason is that when we changed the norm of the space by the function $K(\tau)$ we concentrated only the change of the shape of the unit ball and did not use any correspondence between the points of the two unit balls. Obviously, in a concrete computation we should proceed vice versa, first we should give a correspondence between the points of the old unit ball and the new one and this implies the change of the norm. To this purpose we may define a homotopic mapping $\mathbf{ K }$ which describes the deformation of the norm.
From the lemma above it follows that we can define a homotopic mapping
$$
\mathbf{ K }\left(x,\tau\right): (S,\|\cdot\|_E)\times T \rightarrow (S,\|\cdot\|_E)
$$
such a way that the assumptions:
\begin{itemize}
\item
$\mathbf{ K }\left(x,\tau\right)$ is homogeneous in its first variable and continuously differentiable in its second one,
\item
$ \mathbf{ K }\left(\{e_1,e_2,e_3\},\tau\right)$ is an Auerbach basis of $\left(S,\|\cdot\|^{\tau}\right)$ for every $\tau$,
\item
$ \mathbf{ K }\left(B_E,\tau\right)=K(\tau) $
\end{itemize}
holds.
The mapping $\mathbf{ K }\left(x,\tau\right)$ determines the changes at all levels for example a frame is ``at rest" if its change arises only from this globally determined change, and ``moves with constant velocity" if its origin has this property and the directions of its axes are ``at rest". Precisely, we say, that

\begin{defi}
The frame $\{f_1(\tau),f_2(\tau),f_3(\tau),o(\tau)\}$ \emph{ moves with constant velocity with respect to the time-space}  if for every pairs $\tau$, $\tau'$ in $T^+$ we have
$$
f_i(\tau )=\mathbf{ K }\left(f_i(\tau'),\tau \right) \mbox{ for all } i \mbox{ with } 1\leq i \leq 3
$$
and  there are two vectors $O=o_1e_1+o_2e_2+o_3e_3\in S$ and  $v=v_1e_1+v_2e_2+ v_3e_3 \in S$ that  for all values of $\tau$ we have
$$
O(\tau)=\mathbf{K}(O,\tau)+\tau \mathbf{K}(v,\tau).
$$
A frame is \emph{at rest with respect to the time-space} if the vector $v$ is the zero vector of $S$.
\end{defi}
Consider the derivative of the above equality by $\tau$. We get that
$$
\dot{O}(\tau)=\frac{\partial \mathbf{K}(O,\tau)}{\partial \tau}+ \mathbf{K}(v,\tau)+ \tau \frac{\partial \mathbf{K}(v,\tau)}{\partial \tau},
$$
showing that for such a homotopic mapping, which is constant in the time $O(\tau)$, is a line with direction vector $v$ through the origin of the time space. Similarly in the case when $v$ is the zero vector it is a vertical (parallel to $T$) line-segment through $O$.

\begin{example}
For a simple example (of dimension 3) consider the second example of subsection 3.3. The function $\mathbf{ K }$ can be get by the formula:
$$
\mathbf{ K }\left((x,y)^T, \tau\right)=\left(\alpha x\cos \tau-\frac{1}{\alpha}y \sin \tau, \alpha x \sin \tau+\frac{1}{\alpha}y\cos \tau\right)^T.
$$
Then we have
$$
\mathbf{ K }\left(B_E, \tau\right)=\left(\begin{array}{cc}
\cos \tau & \sin \tau \\
-\sin \tau & \cos \tau
\end{array}\right)G_\alpha
$$
furthermore we get also that
$$
\mathbf{ K }\left(e_1, \tau\right)=\left(\alpha \cos \tau, \alpha \sin \tau\right)^T \mbox{ , }
\mathbf{ K }\left(e_2, \tau\right)=\left(-\frac{1}{\alpha} \sin \tau, \frac{1}{\alpha} \cos \tau\right)^T
$$
gives an Auerbach basis for the corresponding norm.
The unit vectors of a frame at rest can be gotten by the affinity
$$
\left(\begin{array}{cc}
\alpha\cos \tau & \frac{1}{\alpha}\sin \tau \\
-\alpha\sin \tau & \frac{1}{\alpha}\cos \tau
\end{array}\right)
$$
using for the vectors
$$
\left(\cos \beta,\sin \beta \right)^T \mbox{ , }
\left(-\sin \beta , \cos \beta\right)^T,
$$
respectively. (Here $\beta $ is a given parameter.) With respect to the absolute coordinate-system the world-line of the origin is a helical
$$
\tau \mapsto \left(\alpha o_1 \cos \tau + \frac{1}{\alpha}o_2\sin \tau ,-\alpha o_1\sin \tau + \frac{1}{\alpha}o_2\cos \tau \right)^T
$$
through a given point $O=(o_1,o_2)^T$ of the plane $S$.
\end{example}

\subsection{Time-axes}

First we recall a calculation of Subsection 2.3 can be found in present paper before the definition of the product. Consider the unit vector $e\in S$ (with respect to the Euclidean norm) and a two plane generated by the vectors $e$ and $e_4$. This plane intersects the set of light-like vectors in a curve defined by
$$
[\alpha e, \alpha e]^{\tau}+c^2\left[\tau e_4,\tau e_4\right]=0
$$
or  equivalently
$$
 \alpha ^2[e,e]^{\tau}=c^2\tau^2.
$$
From this we get that
$$
\alpha_e (\tau) =\pm \frac{c\tau}{\|e\|^{\tau}}
$$
is the union of two functions of $\tau$ corresponding to the two signs in formula, respectively. If now the sign is positive and we consider a parameter $\beta$ with $|\beta |\leq 1$ the functions
$$
\beta \frac{\tau}{\|e\|^{\tau}}=\beta \alpha_e(\tau)
$$
defines again a set of curves $\tau\rightarrow t_{e,\beta}(\tau)=\alpha_\beta(\tau)e+\tau e_4$ which gives a one-fold covering of the set of time-like points of the corresponding plane.
Natural to say that this  system of curves is a system of \emph{(curvilinear) time axes}. Each of it is a world-line of a particle which velocity vector at the point $\tau$ is
$$
\tau\left(\left(\frac{\beta c}{\|e\|^{\tau}}-\frac{1}{2}\beta c\tau \frac{\frac{\partial [e,e]^{\tau}}{\partial \tau}(\tau)}{\sqrt{[e,e]^{\tau}}\left([e,e]^{\tau}\right)^2}\right)e+ e_4\right).
$$
As we saw in Subsection 2.3 there is no orthogonality at the intersection point of this time-axis and the imaginary unit sphere. Again the product of a tangent vector and the position vector of the point of the intersection is
$$
\frac{[e,\beta\alpha^e(\tau^\star)e]^{\tau^\star}\left(-\beta\frac{\partial{\|\beta\alpha^e(\tau^\star)e\|^{\tau}}} {\partial \tau}(\tau^\star)\right)}{1- \beta\frac{\partial{\|\beta\alpha^e(\tau^\star)e\|^{\tau}}} {\partial \tau}(\tau^\star)},
$$
and so the product of the velocity vector of the time axis with the corresponding tangent vector is
$$
=\frac{-\frac{1}{2}\frac{\partial[e,\beta\alpha^e(\tau^\star)]^{\tau}}{\partial \tau}(\tau^\star)\left((c\tau^\star)^2-\beta^2(c\tau^\star)^2-\frac{1}{2}c\tau^{\star}\beta^2(\alpha^e(\tau^\star))^2\frac{\partial[e,e]^\tau}{\partial \tau}(\tau^\star)\right)}{c\tau^\star -\frac{1}{2}\beta^2\left(\alpha^e(\tau^\star)\right)^2\frac{\partial [e,e]^{\tau}}{\partial{\tau}}(\tau^\star)}.
$$
Using the connection among the values of $\beta $, $\tau^{\star}$ and $\alpha^e(\tau^\star)$ we get that it is zero if and only if
$$
(c\tau^\star)^2-\beta^2(c\tau^\star)^2=1=\frac{1}{2}c\tau^{\star}\beta^2(\alpha^e(\tau^\star))^2\frac{\partial[e,e]^\tau}{\partial \tau}(\tau^\star),
$$
or
$$
1=\frac{1}{2}\frac{(c\tau^\star)^3-c\tau^\star}{[e,e]^{\tau^\star}}\frac{\partial[e,e]^\tau}{\partial \tau}(\tau^\star).
$$
This is a separable differential equation of the function $[e,e]^{\tau}$ with solution
$$
[e,e]^{\tau^\star}=\left(1-\frac{1}{(c\tau^\star)^2}\right)c_e^2
$$
where $c_e$ is a constant depending on the direction $e$. This proves the following lemma:

\begin{lemma}
If the time-dependence of the norm defined by the equalities:
$$
[e,e]^{\tau}=\left(1-\frac{1}{(c\tau)^2}\right)c_e^2 \qquad c_e\in \mathbb{R}
$$
then the imaginary unit sphere and the time-axis $t_{\beta,e}$ intersect to each other orthogonally.
\end{lemma}

The function $\mathbf{ K }$ gives a new chance to define the concept of time-axes. The new definition gives back the concept of $t_{\beta,e}$ if we assume that $\mathbf{ K }$ is invariant on those two-planes, which are defined by the directions of $S$ and the absolute time-axis.

\begin{defi}
The \emph{time-axis} of the time-space model is the world-line $O(\tau)$ of such a particle which moves with constant velocity with respect to the time-space and starts from the origin. More precisely, for the world-line $\left(O(\tau),\tau\right)$ we have $\mathbf{K}(O,\tau)=0$ and hence with a given vector $v\in S$,
$$
O(\tau)=\tau \mathbf{K}(v,\tau).
$$
\end{defi}

\begin{example}
Let the function $\mathbf{ K }$ is defined (as in the previous example) with the equality:
$$
\mathbf{ K }\left((x,y)^T, \tau\right)=\left(\alpha x\cos \tau-\frac{1}{\alpha}y \sin \tau, \alpha x \sin \tau+\frac{1}{\alpha}y\cos \tau\right)^T,
$$
then the time-axis defined by the vector $v=(v_1,v_2)^T$ is the curve
$$
\left(\tau\left(\alpha v_1\cos \tau-\frac{1}{\alpha}v_2 \sin \tau\right),\tau\left( \alpha v_1 \sin \tau+\frac{1}{\alpha}v_2\cos \tau\right),\tau\right)^T.
$$
\end{example}

For the point of intersection of the time-axis and the imaginary sphere of parameter $c$ holds the equality:

$$
(\tau^\star)^2\left(\left[ \mathbf{K}(v,\tau^\star),\mathbf{K}(v,\tau^\star)\right]^{\tau^\star}-c^2\right)=-1
$$
and thus also get that
$$
\left[ \mathbf{K}(v,\tau^\star),\mathbf{K}(v,\tau^\star)\right]^{\tau^\star}=\left(c^2-\frac{1}{(\tau^\star)^2}\right).
$$
We note that for an arbitrary vector $v$ (its unit vector $v^0$) and a parameter $\tau$ we have the equality
$$
\left[ \mathbf{K}(v,\tau),\mathbf{K}(v,\tau)\right]^{\tau}=\|v\|^2_E\left[ \mathbf{K}(v^0,\tau),\mathbf{K}(v^0,\tau)\right]^{\tau}=\|v\|^2_E
$$
simplifying the above formula to the another one
$$
\|v\|^2_E=\left(c^2-\frac{1}{(\tau^\star)^2}\right),
$$
or equivalently
$$
(\tau^\star)^2=\frac{1}{c^2-\|v\|^2_E}.
$$
Now we determine the "angle" between the imaginary unit sphere and the time-axis defined above. The velocity vector of the time-axis at the examined point is
$$
\tau^\star\mathbf{K}(v,\tau^\star)+\left(\tau^\star\right)^2\frac{\partial \mathbf{K}(v,\tau)}{\partial \tau}\left(\tau^\star\right)+ \tau^\star e_4.
$$
If we recalculate the tangent vector of the imaginary unit sphere at its point $s+\tau e_4$ using the opportunity $c(t+\lambda)=s+\lambda e$, we get that it is
$$
\dot{c}(t)+
\frac{2\left[\dot{c}(t),{c}(t)\right]^{\tau}} {2c^2\tau- \frac{\partial \left[c(t),c(t)\right]^{\tau}}{\partial \tau }(\tau)}e_4
$$
The product is
$$
\left[\dot{c}(t),\tau^\star\mathbf{K}(v,\tau^\star)+\left(\tau^\star\right)^2\frac{\partial \mathbf{K}(v,\tau)}{\partial \tau}\left(\tau^\star\right)\right]^{\tau^\star}
-c^2\frac{2\tau^\star\left[\dot{c}(t),{c}(t)\right]^{\tau^\star}}{2\tau^\star  c^2-\frac{\partial \left[c(t),c(t)\right]^{\tau}}{\partial \tau }(\tau^\star)}=
$$
$$
=\left[\dot{c}(t),\tau^\star\mathbf{K}(v,\tau^\star)+\left(\tau^\star\right)^2\frac{\partial \mathbf{K}(v,\tau)}{\partial \tau}\left(\tau^\star\right)\right]^{\tau^\star}-
$$
$$
-\left[\dot{c}(t),\frac{2c^2\tau^\star{c}(t)}{2\tau^\star  c^2-\frac{\partial \left[c(t),c(t)\right]^{\tau}}{\partial \tau }(\tau^\star)}\right]^{\tau^\star}.
$$
Since we have
$$
\tau^\star\mathbf{K}(v,\tau^\star)=s=c(t)
$$
this formula can be simplified into the form
$$
=\left[\dot{c}(t),c(t)+\left(\tau^\star\right)^2\frac{\partial \mathbf{K}(v,\tau)}{\partial \tau}\left(\tau^\star\right)\right]^{\tau^\star}
-\left[\dot{c}(t),\frac{2c^2\tau^\star}{2\tau^\star  c^2-\frac{\partial \left[c(t),c(t)\right]^{\tau}}{\partial \tau }(\tau^\star)}c(t)\right]^{\tau^\star}.
$$
We can see that it is zero in two important cases, the first one is when the function $\mathbf{K}(v,\tau)$ does not depend on the time. The another case is when the following equation system holds with certain function $\alpha(\tau^\star)$:
\begin{eqnarray*}
\frac{\partial \mathbf{K}(v,\tau)}{\partial \tau}\left(\tau^\star\right)&=&\alpha(\tau^\star)c(t)\\
1+(\tau^\star)^2\alpha(\tau^\star)&=&\frac{2c^2\tau^\star}{2\tau^\star  c^2-\frac{\partial \left[c(t),c(t)\right]^{\tau}}{\partial \tau }(\tau^\star)}.
\end{eqnarray*}
This equation system leads to the equality
$$
(\tau^\star)^2\left(2\tau^\star  c^2-\frac{\partial \left[c(t),c(t)\right]^{\tau}}{\partial \tau }(\tau^\star)\right)\frac{\partial \mathbf{K}(v,\tau)}{\partial \tau}\left(\tau^\star\right)
=\frac{\partial \left[c(t),c(t)\right]^{\tau}}{\partial \tau }(\tau^\star)c(t).
$$
For an example define the shape function by the scalar valued function
$$
\mathbf{K}(v,\tau)=\alpha(v,\tau)v.
$$
Then we get that
$$
\frac{\partial \mathbf{K}(v,\tau)}{\partial \tau}=\frac{\partial \alpha(v,\tau)}{\partial \tau}v
$$
and also we have
$$
\mathbf{K}(c(t),\tau)=\alpha(c(t),\tau)c(t),
$$
implying the equality
$$
\alpha^2(c(t),\tau)\left[c(t),c(t)\right]^\tau=\|c(t)\|_E^2.
$$
Since $\alpha(v,\tau)\neq 0$, from
$$
\left[c(t),c(t)\right]^\tau=\frac{\|c(t)\|_E^2}{\alpha^2(c(t),\tau)}
$$
we get that
$$
\frac{\partial \left[c(t),c(t)\right]^{\tau}}{\partial \tau }=-\frac{2\|c(t)\|_E^2}{\alpha^3(c(t),\tau)}\frac{\partial \alpha(c(t),\tau)}{\partial \tau}.
$$
The orthogonality condition for a general $\tau$ means the equality
$$
\tau^2\left(2\tau  c^2+\frac{2\|c(t)\|_E^2}{\alpha^3(c(t),\tau)}\frac{\partial \alpha(c(t),\tau)}{\partial \tau}\right)\frac{\partial \alpha(c(t),\tau)}{\partial \tau}v=-\frac{2\|c(t)\|_E^2}{\alpha^3(c(t),\tau)}\frac{\partial \alpha(c(t),\tau)}{\partial \tau}c(t)
$$
and again if the function $\alpha(v,\tau)$ is a constant we have a solution. In the other case, we can simplify it with its derivative and get that
$$
(\tau)^2\left(2\tau  c^2+\frac{2\|c(t)\|_E^2}{\alpha^3(c(t),\tau)}\right)\frac{\partial \alpha(c(t),\tau)}{\partial \tau}v=-\frac{2\|c(t)\|_E^2}{\alpha^3(c(t),\tau)}c(t).
$$
We also know the connection between $c(t)$ and $v$, because at the point $\tau^\star$ we have
$$
c(t)=\tau^\star\mathbf{K}(v,\tau^\star)=\tau^\star\alpha(v,\tau^\star)v.
$$
This simplifies the above equality to equality among scalar functions:
$$
(\tau)^2\left(2\tau  c^2+\frac{2\|c(t)\|_E^2}{\alpha^3(c(t),\tau)}\right)\frac{\partial \alpha(c(t),\tau)}{\partial \tau}=-\frac{2\|c(t)\|_E^2}{\alpha^3(c(t),\tau)}\tau^\star\alpha(c(t),\tau^\star),
$$
which can be written in the form
$$
-\frac{\tau^3c^2}{\tau^2+\tau^\star\alpha(c(t),\tau^\star)}=\frac{\frac{\partial \alpha(c(t),\tau)}{\partial \tau}}{\alpha^3(c(t),\tau)}.
$$
Solving this separable differential equation, we get that with a constant $C$
$$
\alpha^2(c(t),\tau)=
$$
$$
=\frac{(\tau^\star)^2\alpha^2(c(t),\tau^\star)\|v\|_E^2}{c^2\left(\tau^2-\tau^\star\alpha(c(t),\tau^\star) \ln(\tau^2+\tau^\star\alpha(c(t),\tau^\star))\right)+(\tau^\star)^2\alpha^2(c(t),\tau^\star)\|v\|_E^2C}.
$$
To get the identity at the point $\tau^\star$ we substitute it and so we have
$$
C=\frac{(\tau^\star)^2\left(\|v\|_E^2-c^2\right)+c^2\tau^\star\alpha(c(t),\tau^\star) \ln((\tau^\star)^2+\tau^\star\alpha(c(t),\tau^\star))}{(\tau^\star)^2\alpha^2(c(t),\tau^\star)\|v\|_E^2}.
$$
With this constant the required equality on $\alpha(c(t),\tau)$ is
$$
\alpha^2(c(t),\tau)=\frac{(\tau^\star)^2\alpha^2(c(t),\tau^\star)\|v\|_E^2}{c^2\tau^2-(\tau^\star)^2\left(c^2-\|v\|_E^2\right)
-c^2\tau^\star\alpha(c(t),\tau^\star)\ln\left(\frac{\tau^2+\tau^\star\alpha(c(t),\tau^\star)}{(\tau^\star)^2+ \tau^\star\alpha(c(t),\tau^\star)}\right)}.
$$
The function $\alpha(c(t),\tau)$ is well-defined real valued function if the right hand side is greater or equal to zero. From this assumption we get the equality
$$
\tau^2-\tau^\star\alpha(c(t),\tau^\star)\ln\left(\tau^2+\tau^\star\alpha(c(t),\tau^\star)\right)\geq
$$
$$
\geq\left(1-\frac{\|v\|_E^2}{c^2}\right)(\tau^\star)^2-\tau^\star\alpha(c(t),\tau^\star) \ln\left((\tau^\star)^2+\tau^\star\alpha(c(t),\tau^\star)\right).
$$
Since the left hand side is a monotone increasing function of its variable $\tau\geq 0$, we have to pick up a value in which the equality holds to determine a range interval where this equality also holds. It is easy to calculate that at the value
$$
\tau=\sqrt{\left(1-\frac{\|v\|_E^2}{c^2}\right)}\tau^\star
$$
the equality holds thus $\alpha^2(c(t),\tau)$ can be defined well if $\tau\geq\sqrt{\left(1-\frac{\|v\|_E^2}{c^2}\right)}\tau^\star$.

Using the assumption that the point $c(t)$ is on the imaginary sphere of parameter $c$ we get that
$$
\alpha(c(t),\tau^\star)^2=c^2{\tau^\star}^2-1,
$$
and thus
$$
\alpha^2(c(t),\tau)=
$$
$$
=\frac{(\tau^\star)^2(c^2{\tau^\star}^2-1)\|v\|_E^2}{c^2\tau^2-{\tau^\star}^2\left(c^2-\|v\|_E^2\right)-\tau^\star\sqrt{c^2(\tau^\star)^2-1} \ln\left(\frac{\tau^2+\tau^\star\sqrt{c^2(\tau^\star)^2-1}}{(\tau^\star)^2+\tau^\star \sqrt{c^2(\tau^\star)^2-1}}\right)}.
$$

\subsection{Remark on cosmology}

Our model can be considered also as a new model of the universe. The deterministic variant obviously contains as a special case the model of Minkowski space-time. On the other hand it can be extended to a generalization of the Robertson-Walker space-time, too. (To this we have to change by the time the volume of the unit ball of the space-like subspace $S$ and we have to allow it one of the metric of the three spaces of constant curvature. From the Minkowski product these metrics can be defined without any difficulties.) The advantage of our model that $S$ can be considered also as a general normed space (without inner product).

The deterministic time-space can be considered in a non-deterministic way, too. Thus we gave a concept of random time-space and proved that (on a finite range of time) every such space can be approximated with a deterministic model well. (In this section we assume that the volume of the unit ball does not depend on the time but this condition can be omitted in the rest of this paper.)

The time-space can also be defined in a more convenient way, using a shape function. It regulates the methods of calculations in time-space and gives the possibility to rewrite the equality of special and global relativity.

\begin{center}
\'Akos G. Horv\'ath,\\
 Department of Geometry \\
 Mathematical Institute \\
Budapest University of Technology and Economics\\
1521 Budapest, Hungary
\\e-mail: ghorvath@math.bme.hu
\end{center}

\end{document}